# Orthogonality between cellulose nanocrystals and a low-molecular weight gelator

*Thuy-Linh Phi [a,b], Eero Kontturi [b,\*], Niki Baccile [a,\*]*

[a] Sorbonne Université, Centre National de la Recherche Scientifique, Laboratoire de Chimie de la Matière Condensée de Paris, LCMCP, F-75005 Paris, France

[b] Department of Bioproducts and Biosystems, School of Chemical Engineering, Aalto University, P.O. Box 16300, 00076 Aalto, Finland

**Corresponding Authors**
E-mail: niki.baccile@sorbonne-universite.fr
E-mail: eero.kontturi@aalto.fi

**Keywords**

Cellulose nanocrystals (CNCs) ; Biosurfactants ; Orthogonal hydrogels ; Low molecular weight gelators (LMWGs)

**Abstract**

The development of multicomponent hydrogels has gained a lot of attention in the field of soft matter, as precise tuning of the chemical nature and colloidal properties of each component brings mechanical and functional benefits compared to one-component gels. Within the field, orthogonality between a self-assembled low-molecular weight gelator (LMWG) and a colloid is a domain that has received little attention. In this study, orthogonal LMWG-colloid hydrogels were developed with the additional constraint of sustainability: a bolaamphiphile glycolipid (G-C18:1) is selected as LMWG while cellulose nanocrystals (CNCs) as colloid. These compounds are chosen for their dual role. G-C18:1 is a LMWG but it can also be used, at lower concentrations, as surface stabilizer for CNCs and tune its aggregative properties. On the other hand, tuning surface properties of CNCs drives its bulk behavior: uncharged CNCs locally aggregate and act as reinforcing agent for the LMWG gel, while negatively-charged CNCs, cross-linked with $Ca^{2+}$, naturally form a hydrogel, which can interpenetrate with the LMWG network. By means of rheometry, small-angle X-ray scattering (SAXS) and rheo-SAXS, it is shown here how the aggregative behavior of CNCs enhances the mechanical properties of G-C18:1 hydrogels, while G-C18:1 imparts pH and temperature responsiveness to CNC hydrogels.

**Introduction**

Hydrogels, three-dimensional cross-linked or entangled networks typically composed of synthetic or natural macromolecules possess the remarkable ability to retain significant volumes of water.[1,2] This characteristic renders them versatile for applications across various fields, including biomedicine[3,4] and the food industry,[5,6] where their biocompatibility, biodegradability, and unique rheological features are necessary. In addition to the typical building blocks, hydrogels can also be manufactured, for example, from low molecular weight gelators (LMWGs). The purpose of this paper is to explore a particular case of orthogonal bicomponent hydrogels made of LMWGs and polymeric nanoparticles called cellulose nanocrystals (CNCs) with a special emphasis on their responsive properties.

One strategy to improve properties of hydrogels is the development of multicomponent gels. In the absence of cross-linking, they can be categorized in various types: interpenetrated polymer networks (IPNs), involve two or more covalently cross-linked polymer networks (or polymer-LMWG systems) that are physically entangled but not chemically bonded;[7] self-sorting, commonly formed by LMWGs, consist of distinct networks where each gelator self-assembles independently into separate fibrous structures, following its own supramolecular pathway;[8] orthogonal, generally referring to *"the independent formation within a single system of different supramolecular structures, each with their own characteristics"*.[9] Orthogonal gels generally involve two independent components, where one may contribute to the hydrogel by forming fibrous structures, for instance a fibrous gel network from a LMWG or peptide and a secondary structure (micelles, vesicles) obtained by a surfactant.[9,10] Polymeric orthogonal gels are also well-discussed in the literature.[11] Interestingly, orthogonal hydrogels are studied since long time,[12] with obvious applications in the field of biomedicine and drug delivery.[13]

Among all possible systems, LMWGs have garnered attention for their capacity to form self-assembled fibrillar networks (SAFiNs) in aqueous environments.[14] Driven by non-covalent interactions into three-dimensional isotropic self-assembled structures, LMWGs are naturally conceived as stimuli-responsive supramolecular systems,[15,16] with reversible gelation, low concentration requirements, minimal use of cross-linkers, and simple preparation.[17] Consequently, developing strategies to construct SAFiNs-based hydrogels with orthogonal architectures by combining LMWGs with other components, such as polymer,[7] surfactants,[18] biobased macromolecules[19,20] or other LMWGs,[21] has attracted considerable attention. Nevertheless, few research pathways have explored the orthogonality between LMGWs and nanosized colloids, the latter with tunable aggregation properties, and adding the constraint of sustainability.

An interesting field of research in soft matter science is the development of orthogonal hydrogels containing cellulose and LMWGs, although, to the best of our knowledge, there are no existing reports. Since cellulose is among the most extensively studied macromolecules in the field of soft matter, particularly for biomedical applications,[22,23] developing and studying the properties of orthogonal hydrogels containing CNCs and a LMWG represents an intriguing avenue for research, and this for two reasons. First of all, CNCs are bio-based nanoparticles, relevant for the development of sustainable nanoscale science and engineering. Secondly, the surface chemistry of CNCs can be controlled in such a way to tune their aggregation and dispersion properties, making them interesting either as reinforcements in hydrogels, [24,25] or as hydrogel scaffold themselves.[26,27] These aspects were never explored in the context of orthogonal hydrogels.

In this work, we then study orthogonality in fully bio-based hydrogels composed of a single glucose lipid LMWG (G-C18:1) and CNCs. G-C18:1 is selected for its multiphasic behavior in water at room temperature[28] and linked to its unique surfactant-lipid-gelator nature,[28,29] tuned by pH and/or type of ion. Below neutral pH and at concentrations under 5 wt%, G-C18:1 forms vesicles, displaying a lipid-like behavior. At pH above neutrality, it assembles into micelles, thus exhibiting a surfactant behavior.[30,31] When $Ca^{2+}$ is added to its micellar phase,[32,33] G-C18:1 forms fiber gels[30] (Figure 1). In particular, we focused on orthogonal G-C18:1/CNC hydrogels, in which CNCs either assembled into hydrogels (negatively-charged and cross-linked by calcium ions, referred to as SCNCs, Figure 1) or behaved as reinforcing agents (uncharged CNCs prepared via HCl hydrolysis, referred to as $CNC_\alpha$, and neutral surface stabilized by G-C18:1,[34] Figure 1). These two types of CNCs exhibit distinct roles in the hydrogel system: SCNCs actively participate in the formation of a percolated network, while $CNC_\alpha$ are used to reinforce the hydrogel matrix physically. Specific attention is paid to the impact of the assembled form of CNCs to the elastic properties of the LMWG hydrogel as well as how the responsivity to pH and temperature of the LMWG affect the elastic properties of CNC hydrogels.[33]

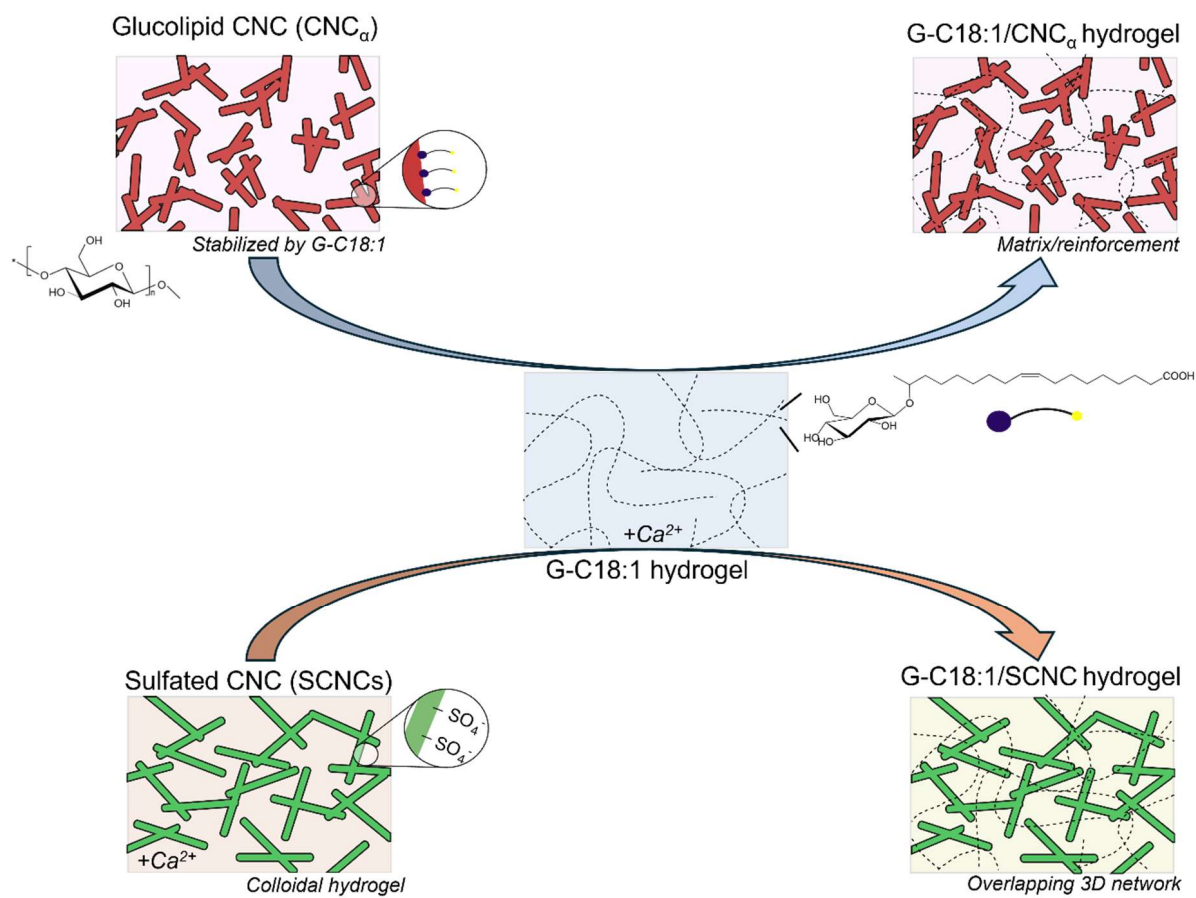

**Figure 1.** Scheme illustrating the strategy to prepare G-C18:1/CNC multicomponent hydrogels. Two sources of CNCs are chosen, colloidal hydrogels of calcium-containing sulfated CNCs (SCNCs) and colloidal suspension of uncharged CNC stabilized by G-C18:1 (CNC$_α$, $α$= 0.2 or 1, refer to the Materials and Method section).[34] G-C18:1 assembles in a fibrous self-assembled hydrogel in the presence of a calcium source.

**Materials and Methods**

**Chemicals.** Monounsaturated glucolipid G-C18:1 (Mw= 462 g/mol) is described in previous studies[28,31] and obtained from Amphistar in Gent, Belgium, and produced at the Bio Base Europe Pilot Plant in Gent, Belgium, under batch No. APS F06/F07, Inv96/98/99. This compound was employed as such. To adjust the pH, microliter amount (0.1 M, 0.5 and 1 M) of NaOH and HCl solutions (Sigma-Aldrich, Germany) were used. The gelation of G-C18:1 was trigged by adding microliter amounts of a 1 M CaCl$_2$ (VWR) solution. Whatman 1 filter papers (Catalog No. 1001 150), were used as the cellulose substrate. According to the manufacturer, filter papers are primarily composed of purified cotton linters, with a cellulose content exceeding 98%. Aqueous hydrochloric acid (~37% concentration, Sigma-Aldrich) was used from a concentrated stock solution. High sulfur content cellulose nanocrystals prepared from H$_2$SO$_4$ hydrolysis (**SCNCs**), with widths of 10–20 nm and lengths of 300–900 nm, were obtained from Nanografi (Turkey). For all dilution, washing, and rinsing steps, Milli-Q water was used.

**Hydrolyzed cellulosic (HC) fibers: preparation method.** The preparation of hydrolyzed cellulosic fibers was performed using the aqueous HCl approach in a conventional liquid-solid system followed the method described by Rusli *et al.*[35] To start, 10 grams of Whatman 1 filter paper were mixed with 300 milliliters of pre-heated 3 M hydrochloric acid (HCl) solution, with continuous stirring at 1500 rpm at 80°C for 4 hours. The resulting suspension was then subjected to multiple rounds of centrifugation at 1460 g (2900 rpm) for 10 minutes each, until the pH reached between 4 and 5. Afterward, the solution was dialyzed for three days using a Spectra/Por® Dialysis Membrane (molecular weight cutoff 6-8 kD, part number 132665).

**Preparation method of $CNC_\alpha$ samples (Figure 2a): dispersion of CNCs from hydrolyzed cellulosic fibers by G-C18:1.** In order to obtain CNCs from hydrolyzed cellulosic (HC) fibers, which otherwise lack sufficient surface charge to prevent aggregation,[36,37] it is necessary to implement an additional stabilizing step. On the basis of our previous work,[34] we used small amounts of G-C18:1 to disperse the CNCs. First, a 10 mL solution of HC fibers at 4 wt% (mass of HC= 0.4 g) was prepared and subjected to tip sonication (UP200Ht, Hielscher) at 100 W power and 20% amplitude for 10 minutes. Subsequently, specific amounts of G-C18:1 powder were added directly to the HC fiber medium to obtain a CNCs suspension. The resulting samples are labeled $CNC_\alpha$, with α= 0.2 or 1. More specifically, sample $CNC_{0.2}$ refers to a normalized mass fraction of CNC:G-C18:1 equal to 1:0.2 (0.08 g of G-C18:1), while $CNC_1$ refers to a normalized mass fraction of 1:1 (0.4 g of G-C18:1) (Figure 2a). The mixtures were tip-sonicated again at 10% amplitude for 5 min until a homogeneous solution was achieved. This protocol was developed by us in a recent work.[34]

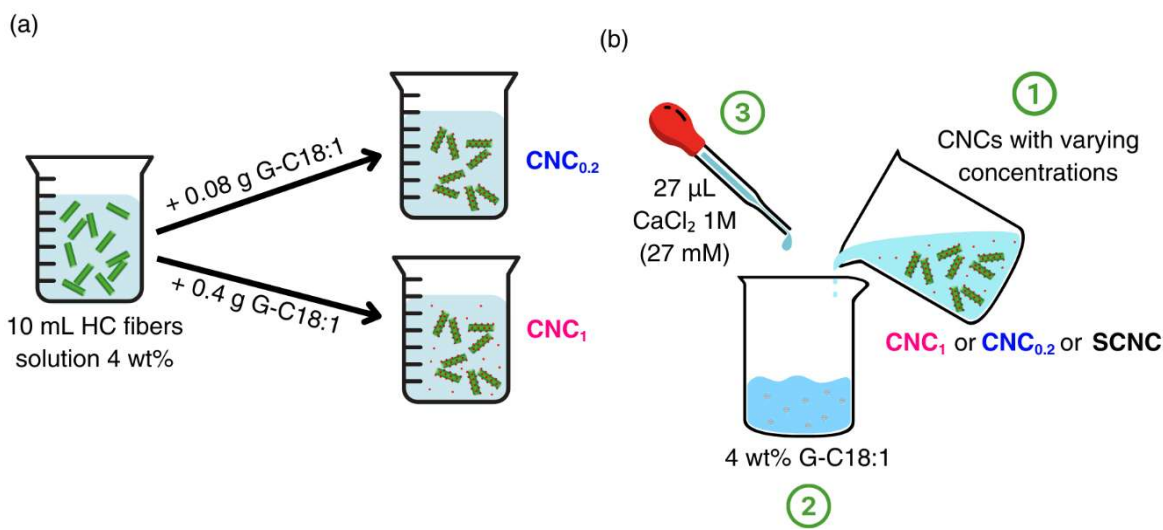

Figure 2. (a) Schematic illustration of the production of a typical colloidal suspension of uncharged CNC, referred in the manuscript as $CNC_\alpha$, with α= 0.2 or 1, as explained below. $CNC_\alpha$ samples are generated *in situ* from hydrolyzed cellulose (HC) fibers. The uncharged cellulose nanocrystals are surface stabilized using small amounts of the glucolipid G-C18:1 based on our previous study.[34] In brief, a 10 mL solution of HC fibers (0.4 g, that is a weight fraction of 4 wt%) are tip-sonicated with either 0.08 g or 0.4 g of G-C18:1. The samples are then labeled as $CNC_{0.2}$ (normalized mass fraction CNC/G-C18:1= 1:0.2) and $CNC_1$ (normalized mass fraction 1:1), respectively. (b) Schematic representation of the method to prepare multicomponent G-C18:1/CNC hydrogels using either the colloidal suspension of uncharged CNC aggregates ($CNC_\alpha$ samples) or sulfated SCNC samples.

**Gels preparation (Figure 2b).** The CNC samples, uncharged colloids ($CNC_\alpha$, with α= 0.2 or 1) or sulfated CNC (**SCNC**), were employed to prepare orthogonal hydrogels with G-C18:1. The preparation of samples $CNC_{0.2}$ and $CNC_1$ was described in the previous section, while **SCNC** was purchased as such. A stock micellar solution of G-C18:1 at 4 wt% was prepared in Milli-Q water using vortexing and sonication. The pH was adjusted to 8.3–8.5 by adding microliter amounts of NaOH solutions (1 M, 0.5 M, and 0.1 M). $CNC_{0.2}$, $CNC_1$ or **SCNC** suspensions were added at various concentrations to the G-C18:1 micellar solution to achieve G-C18:1/CNCs mass ratio ranging from 1:0.05 to 1:2, with a total volume of 1 mL (Table 1 and Table 2). Gelation was initiated by adding 27 μL of 1 M $CaCl_2$ (27 mM), resulting in a $[CaCl_2]/[G\text{-}C18:1]$ molar ratio of 0.5, according to a published protocol.[33] Immediately after the addition of $Ca^{2+}$, the solution was stirred for about 30 s, and gelation began within minutes (Figure 2b). Consistent gel quality, preventing aggregation, heterogeneity and loss of rheological properties (Figure S1) also requires sonication of the G-C18:1 solution before each use.

**Table 1. Preparation of G-C18:1/CNCα hydrogels.** The table provides the solution volumes of G-C18:1 (C= 4 wt%), CNCα (C= 4 wt%) and water to prepare the hydrogels. 27 μL of 1 M $CaCl_2$ (27 mM) are added to each solution (1 mL total volume). The method to prepare CNCα (α= 0.2 or 1) samples is provided in Figure 2a and corresponding section.

| G-C18:1/CNCα mass ratio | Volume G-C18:1 solution / μL | Volume CNCα suspension / μL | Volume $H_2O$ / μL |
|---|---|---|---|
| 1:0.05 | 486.5 | 24.3 | 462.2 |
| 1:0.25 | 486.5 | 121.6 | 364.9 |
| 1:0.5 | 486.5 | 243.3 | 243.3 |
| 1:1 | 486.5 | 486.5 | 0 |

**Table 2. Preparation of G-C18:1/SCNC hydrogels.** Solution volumes of G-C18:1 (C= 4 wt%), SCNCs (C= 8 wt%) and water to prepare studied hydrogels. 27 μL of 1 M $CaCl_2$ (27 mM) are added to each solution (1 mL total volume). SCNC samples are purchased and used as such.

| G-C18:1/SCNCs mass ratio | Volume G-C18:1 solution / μL | Volume SCNCs suspension / μL | Volume $H_2O$ / μL |
|---|---|---|---|
| 1:0.05 | 486.5 | 12.2 | 474.3 |
| 1:0.25 | 486.5 | 30.4 | 456.1 |
| 1:0.5 | 486.5 | 121.6 | 364.9 |
| 1:1 | 486.5 | 243.3 | 243.3 |
| 1:2 | 486.5 | 486.5 | 0 |

**Rheology.** An MCR 302 rheometer (Anton Paar, Graz, Austria) with sand-blasted plate-plate geometry (Ø: 25 mm) was used and maintained at a constant temperature of 25°C. A solvent trap containing water was employed to reduce evaporation. Approximately 0.5 mL of gel was carefully placed in the center of the plate with a spatula to avoid trapping air bubbles, and any excess gel was meticulously removed. All samples were prepared 7 days prior to measurement, unless otherwise specified. The pseudo-equilibrium G′ value was recorded after 5 minutes of oscillatory testing at 1 Hz and low strain ($\gamma$), which was one order of magnitude lower than the critical strain of 0.1%. Frequency sweep experiments were conducted at a strain of 0.1% to be in the linear viscoelastic regime.

**Rheo-SAXS.** Small-Angle X-ray Scattering (SAXS) experiments combined with rheology (Rheo-SAXS) were conducted at the SWING beamline at Synchrotron Soleil, Saint-Aubain, France (Proposal No. 20231446). The beamline operates at an energy of E = 12 keV, with sample-to-detector distances fixed at 2 m and 6 m. The raw data from the 2D detector are azimuthally integrated using the Foxtrot software available at the beamline to generate the typical scattered intensity I(q) profile. Here, q is the wavevector, defined as q = 4π/λ sin(θ), where $2\theta$ is the scattering angle and $\lambda$ is the wavelength. Each frame corresponds to a 500 ms exposure time followed by a 9500 ms resting period, resulting in one frame being captured every 10 seconds. A MCR 501 rheometer (Anton Paar, Graz, Austria) equipped with a Couette polycarbonate cell (10 mm diameter, 0.5 mm gap, V = 1.35 mL) was coupled to the beamline and controlled remotely from an external computer in the experimental hutch using Rheoplus/32V3.62 software. The experiments were conducted at 25 °C in both radial and tangential configurations. In the radial position, the incident beam is perpendicular to the streamlines (also known as velocity lines). In the tangential position, the incident beam passes through the Couette cell parallel to the streamlines, following the direction of flow. Rheology and SAXS data collection were manually synchronized, with a time synchronization error of less than 3 s.

**Results and discussions**

*$CNC_\alpha$ samples as mechanical reinforcing agents for G-C18:1 hydrogels*

The hydrogel properties of G-C18:1, triggered by calcium ions and studied in a set of dedicated works,[28,32,33,38] are reproduced here at a fixed concentration of 2 wt% G-C18:1. The gelation of G-C18:1 is triggered at pH above 7 by adding a source of calcium ions. More details about the rheological properties of G-C18:1 with increasing content of $CaCl_2$ (10, 20 and 30 mM) and monitored in time (from 1 to 60 days) are given in Figure S2, with explanatory text provided in the Supporting Information. Throughout this work, we employ the same conditions, namely a $CaCl_2$ molar concentration of 27 mM and an aging time of 7 days.

Hereafter, we then focus on the development of multicomponent hydrogels involving G-C18:1 gels as matrix and colloidal dispersions of $CNC_\alpha$ samples used as reinforcing agents to implement the mechanical properties of G-C18:1 hydrogels. As discussed in a previous work[34] and in the Materials and Method section, $CNC_\alpha$ ($\alpha$= 0.2 or 1, that is **$CNC_{0.2}$** and **$CNC_1$**) identifies uncharged CNC samples dispersed in water and of which the surface is fully stabilized by the same glucolipid (G-C18:1) employed here to prepare the hydrogel matrix. In this part of the work, G-C18:1 has then a dual role: surface stabilizer for CNCs and gelator. The subscript $\alpha$ then refers to the content of G-C18:1 employed as surface stabilizer, as shown in Figure 2a. On the other hand, the preparation method for the multicomponent gels is illustrated in Figure 2b and the mass ratio between $CNC_\alpha$ and G-C18:1 are given in Table 1.

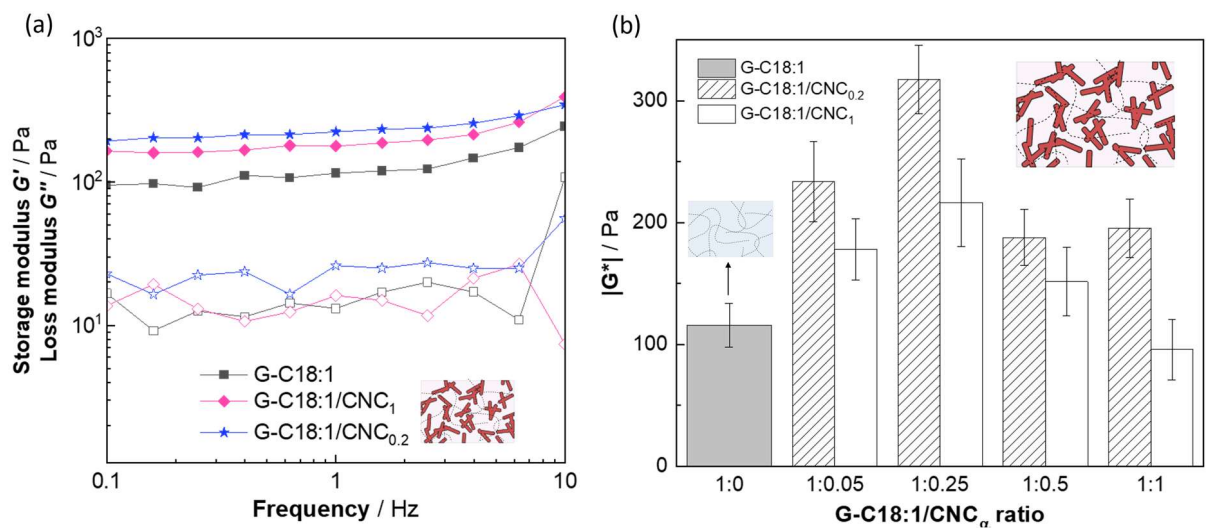

**Figure 3.** Rheological study of G-C18:1 hydrogel reinforced with CNC colloidal aggregates, $CNC_\alpha$, $\alpha = 0.2$ or 1. (a) Frequency sweep measurements show the storage $G'$ (full symbols) and loss $G''$ (empty symbols) moduli for 2 wt% G-C18:1 gels without and with the presence of $CNC_1$ and $CNC_{0.2}$. The G-C18:1/$CNC_\alpha$ is 1:0.05 (Table 1). (b) The complex shear modulus $|G^*|$ of G-C18:1/$CNC_\alpha$ gels given as a function of the ratio between G-C18:1 matrix (2 wt%) and CNC sample.

Figure 3a displays the gel-like, frequency-independent, behavior of $G'$ and $G''$ ($G' > 10G''$) for a G-C18:1 hydrogel and two $CNC_\alpha$-reinforced G-C18:1 hydrogels, prepared with 2 wt% of G-C18:1 and 0.1 wt% CNCs, equivalent to a mass ratio of G-C18:1/CNCs of 1:0.05 (Table 1). Notably, G-C18:1/$CNC_\alpha$ hydrogel samples displayed higher $G'$ values than the CNCs-free control ($G'(1Hz)= 115$ Pa, Figure S2 in the Supporting Information), despite the small mass fraction of $CNC_\alpha$ (1:0.05). This is emblematic of the role of CNCs as reinforcing agents for the hydrogels. On the other hand, the strength of hydrogel G-C18:1/$CNC_{0.2}$ ($G'(1$ Hz$)= 232$ Pa) seems to be higher compared to G-C18:1/$CNC_1$ ($G'(1$ Hz$)= 177$ Pa). The higher $G'$ for G-C18:1/$CNC_{0.2}$ suggests that the amount of G-C18:1 used as surface stabilizer for CNCs also plays an important role. This is not unexpected, as it was shown that the colloidal stability of CNCs in water strongly depends on the content of G-C18:1.[34] G-C18:1 hydrogels were subsequently analyzed with higher CNCs loadings (0.5, 1, and 2 wt%), corresponding to 25%, 50%, and 100% of the dry weight of the biosurfactant. The oscillatory strain measurements at 1 Hz for G-C18:1/$CNC_\alpha$ gels at different mass ratios are presented in Figure S3, while the key hydrogel properties, $G'$ and loss factor (tan δ), were evaluated through small amplitude oscillatory shear measurements and summarized in Table 3.

**Table 3.** Storage modulus (G′) in Pa and loss factor (tan δ = G″/ G′) were measured for G-C18:1/CNC$_\alpha$ where α= 0.2 or 1 at various CNCs concentrations after 7 days. G′= 115.4 Pa and tan δ= 0.08 for the control CNCs-free G-C18:1 hydrogel. Table 1 provides details about the composition of the multicomponent gels.

| Sample | G-C18:1/CNC$_\alpha$ mass ratio | | | | | | | | | |
|---|---|---|---|---|---|---|---|---|---|---|
| | 1:0 | | 1:0.05 | | 1:0.25 | | 1:0.5 | | 1:1 | |
| | G′ | tan δ | G′ | tan δ | G′ | tan δ | G′ | tan δ | G′ | tan δ |
| G-C18:1/CNC$_{0.2}$ | 115.4 | 0.08 | 232.8 | 0.1 | 316.4 | 0.09 | 186.9 | 0.10 | 194.4 | 0.11 |
| G-C18:1/CNC$_1$ | 115.4 | 0.08 | 177.3 | 0.1 | 215.5 | 0.09 | 150.4 | 0.13 | 94.7 | 0.15 |

As shown in Table 3, all systems are hydrogels with tan δ in the order of 0.1. Furthermore, G′ was consistently higher for G-C18:1/CNC$_{0.2}$ compared to G-C18:1/CNC$_1$ across all tested G-C18:1/CNC$_\alpha$ mass ratios. Specifically, G′ of G-C18:1/CNC$_{0.2}$ exceeded that of G-C18:1/CNC$_1$ by a factor ranging from 1.3 to 2.1 within the analyzed concentration range, a sign of a higher resistance to deformation under stress compared to the gel formed with G-C18:1/CNC$_1$. This is also shown by the corresponding values of tan δ, which are consistently low (0.09–0.11), indicating stable elastic dominance regardless of the CNCs concentration.

The complex shear moduli (G*) of all samples, measured at 1 Hz, are plotted in Figure 3b to further compare their viscoelasticity. The complex shear modulus, defined as $G^* = G' + iG''$, reflects the rigidity of a gel under deformation below its yield stress. The general trend is that CNCs reinforcement produces stronger gels compared to the control. In particular, at the G-C18:1/CNC$_\alpha$ mass ratio of 1:0.25 (Figure 3b), the maximum |G*| value for sample G-C18:1/CNC$_{0.2}$ was three times larger than for the control G-C18:1 gel, demonstrating the beneficial reinforcement effect of CNCs. This improvement falls within the typical range observed for hydrogel systems reinforced with CNCs,[24,25] although it remains modest compared to the substantial 2800-fold increase in G′ observed in cellulose micro/nanocrystals reinforced polyurethane system.[39] Furthermore, G-C18:1/CNC$_{0.2}$ samples consistently exhibit higher |G*| values compared to G-C18:1/CNC$_1$, explained by the better colloidal stability of CNC$_{0.2}$, as discussed previously.[34] While the optimal matrix-to-reinforcement ratio was determined to be 1:0.25, hydrogel formation was still achievable at a mass ratio up to 1:1 (Figure S4). However, exceeding the optimal G-C18:1/CNC$_\alpha$ mass ratio of 1:0.25 resulted in a decrease in gel rigidity, indicating that excessive reinforcement can disrupt the network structure.

To better understand the gel network structure, SAXS was employed to analyze the structural organization of G-C18:1/CNC$_{0.2}$ (mass ratio of: 1:0.05, 1:0.25, and 1:1) in comparison to the CNCs and G-C18:1 controls. SAXS measurements of G-C18:1/CNC$_{0.2}$ at ratio 1:0.05 and 1:0.25 (Figure S5 and corresponding discussion in the Supporting Information) shows that the presence of local aggregates of CNCs neither affects the fibers morphology nor their order. The moderate CNCs incorporation effectively reinforces and preserves the G-C18:1 network. On the other hand, the loss in mechanical strength at high CNCs concentrations (1:1 mass ratio) correlates well with the more and more dominant scattering of CNCs in the SAXS profile of mixed gel (see Supporting Information Figure S5). A similar phenomenon has been observed in different nanocomposite systems where CNCs act as reinforcing agent.[40,41] Overall, these findings highlight the importance of maintaining an optimal CNCs loading to balance mechanical reinforcement and structural integrity of the hydrogel network.

Following the experimental line presented in our previous study,[34] we have repeated the rheology and SAXS experiments using a set of HC fibers prepared by gaseous hydrolysis[42] but still stabilized with small content of G-C18:1.[34] The corresponding CNCs, called gCNCs, were then used as reinforcements in G-C18:1 hydrogels. The detailed information regarding this protocol, along with corresponding rheometry and SAXS data, are presented in Figure S6, Table S1 and the accompanying text. Overall, both the mechanical behavior and structural features are very similar to the samples prepared from liquid hydrolysis. It is worth noting, for instance, that the optimum $|G*|$ is observed for the G-C18:1/gCNC$_{0.2}$ mass ratio of 1:0.25 (Figure S6b), as found for CNC$_{0.2}$ (Figure 3b). The absence of notable differences in rheological behavior between hydrogels prepared using aqueous and gaseous CNCs suggests that the hydrolysis method does not significantly influence the final gel properties.

*SCNCs as orthogonal gel network within the G-C18:1 hydrogel matrix*

As previously mentioned, G-C18:1 is not the only component exhibiting the salt ion-induced hydrogel phenomenon. CNCs, particularly sulfated-CNCs (referred to as SCNCs in this study), undergo a sol-to-gel transition either at concentrations above 10 wt%[43] or by cross-linking SCNCs with multivalent ions, as reported in previous studies.[44,45] Studying the rheological, structural and functional properties of an interpenetrated network of SCNCs and a low-molecular weight gelator (LMWG) goes then much beyond the reinforcement functionality of CNCs studied in the previous section. Furthermore, to the best of our knowledge, orthogonal gel constituted by a CNC hydrogel (here, Ca$^{2+}$-cross-linked SCNCs) and a LMWG was never studied before.

A comprehensive assessment of the hydrogel formation process of SCNCs was conducted for concentrations ranging from 0.05 wt% to 2 wt% with 27 mM of CaCl$_2$, as shown in Figure 4a: hydrogels were observed at SCNCs concentrations above 1 wt%. Considering that the same content (27 mM) of $CaCl_2$ triggers hydrogel-formation for both G-C18:1 (2 wt%) and SCNCs (1 wt%), it becomes necessary to explore the optimum amount of $Ca^{2+}$ in the orthogonal G-C18:1/SCNCs system, as the amount of 27 mM may not be sufficient to fully induce gelification in both systems, as observed before for G-C18:1/alginate orthogonal gels.[19]

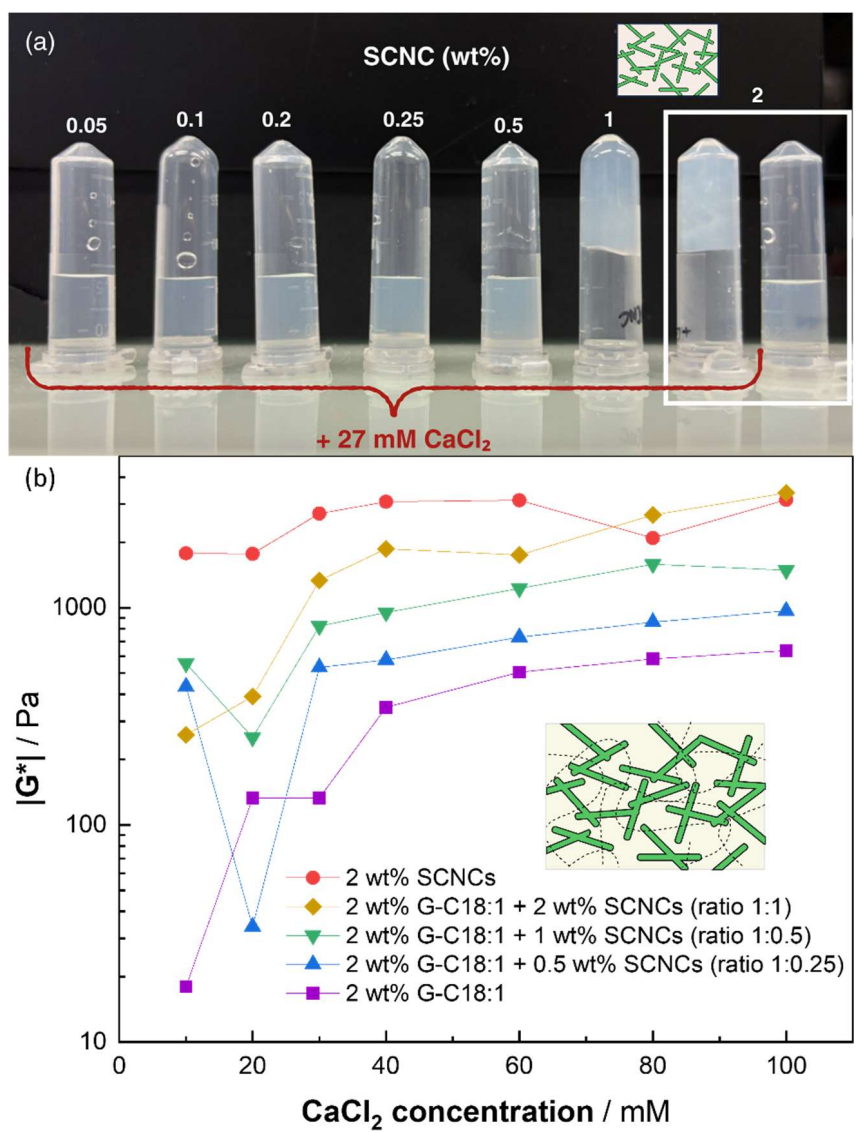

**Figure 4. Rheological study of SCNC hydrogels. (a) Image of SCNC samples at varying concentrations (0.05 to 2 wt%) in the presence of 27 mM $CaCl_2$. The final sample, containing 2 wt% SCNC without $CaCl_2$, serves as a comparison, confirming that gelification occurs due to the presence of $Ca^{2+}$ ions. (b) The complex shear modulus (G*) as a function of $CaCl_2$ concentration (10 to 100 mM) at 1 Hz for 2 wt% G-C18:1, 2 wt% SCNCs, and G-C18:1/SCNC hydrogels. In the orthogonal G-C18:1/SCNC systems, the G-C18:1 concentration is fixed at 2 wt%, while the SCNCs concentration varies according to the mass ratio between G-C18:1 and SCNCs.**

All data for G* as a function of frequency has been systematically analyzed and presented in Figure S7, with the G*(1 Hz) plotted in Figure 4b. These data first of all confirm that the G-C18:1 alone (violet line) exhibits a distinct threshold-type response to $CaCl_2$.[28] Above ~20 mM, G* gradually increases from 133 Pa to 636 Pa. In contrast, hydrogels prepared from 2 wt% SCNCs only (red curve) exhibit significantly higher G* values, approximately an order of magnitude higher than those of the pure G-C18:1 gels, even at $CaCl_2$ concentrations below 20 mM. This suggests that SCNCs form a percolated, rigid network at relatively low salt levels and that G* does not exhibit an actual dependence on $CaCl_2$ content (G* fluctuates between 2000 and 3000 Pa), at least in the regime of calcium content explored here. These observations agree with the findings reported by Lu *et al.*, on salt-ion effects in SCNCs hydrogels.[44] When G-C18:1 and SCNCs are combined, their interactions vary depending on the SCNCs content. At SCNCs loading (1:0.25 ratio, blue curve) below its critical gelation concentration of 1 wt%, the gel strength is primarily governed by G-C18:1, with G* following a similar trend to that of the pure G-C18:1 hydrogels. At this stage, the increase in G* occurs above 20 mM $CaCl_2$, that is, the critical gelation point of G-C18:1. Above 20 mM $CaCl_2$, G* reaches higher values in the orthogonal gels than in the G-C18:1 monocomponent gels, nicely indicating that SCNCs actively improves the elastic properties of G-C18:1 gels, probably acting as a reinforcing agent, as observed for $CNC_\alpha$ systems.

G* systematically increases at higher SCNCs concentration and for calcium content above 20 mM, reaching values close to the SCNCs control, yet slightly below, probably due to the partitioning of calcium ions between the SCNCs and G-C18:1 networks. If cross-component linking with an effect on the overall mechanical properties is a mechanism that cannot be excluded, the fact that the highest G* never exceed the G* of SCNCs could be an indication that this phenomenon either does not occur or it is not rheologically relevant. Indeed, cross-linking in supramolecular hydrogels was shown to enhance the mechanical properties.[46,47]

These results suggest a change in the role of the CNCs from the first to the second part of this work. If $CNC_\alpha$ behaves as reinforcement, SCNCs at high concentration (starting at 1 wt%) rather behave as a co-network in the orthogonal gels above the 20 mM threshold. Only the gel behavior at 20 mM $Ca^{2+}$, the critical gelation point of G-C18:1, raises some questions: the value of G* at 20 mM in the orthogonal gels (Figure 4b) drops for the G-C18:1/SCNC 1:025 (blue curve) and 1:0.5 (green curve) ratio, while it recovers with the increasing trend at 1:1 (yellow curve). A plausible explanation could be that, at borderline $Ca^{2+}$ concentrations, calcium ions partition in a non-homogeneous fashion between the SCNCs and G-C18:1 components, thereby hindering the complete crosslinking of G-C18:1, as also observed in the case of G-C18:1/alginate hydrogels.[19] At a 1:1 ratio (Figure 4b, yellow curve), where SCNCs gelation predominates, the drop in G* at 20 mM CaCl$_2$ is no longer observed, supporting this hypothesis. SAXS (Figure S8 and related discussion in the Supporting Information), used in this system as well to probe the local structure of G-C18:1/SCNC hydrogels, also confirms the overlapping 3D structure of G-C18:1 and SCNCs networks. Overall, these findings suggest that the addition of 27 mM of Ca$^{2+}$ ions in the G-C18:1/SCNCs system is beyond the co-gelation threshold, thus successfully facilitating the formation of two overlapping, three-dimensional networks comprising both G-C18:1 and SCNCs.

*Rheo-SAXS of G-C18:1/CNC$_{0.2}$ and G-C18:1/SCNC gels*

To better understand the rheological properties of orthogonal gels in relationship to their structure, we studied the thixotropic behavior of G-C18:1/CNC$_{0.2}$ and G-C18:1/SCNC hydrogels with a 1:1 ratio by means of small-angle X-ray scattering (SAXS) combined with rheometry (rheo-SAXS), as shown in Figure 5. The plateau value of G′ sets at approximately 750 Pa for G-C18:1/CNC$_{0.2}$ and 800 Pa for G-C18:1/SCNC (Figure 5a,b, respectively), with the absolute values differing from those reported earlier, possibly due to the use of different measurement geometries (sand-blasted plate-plate versus concentric cylinders – Couette cells), as the calculation of rheological parameters depends on the geometry used.[48]

A preliminary strain-sweep test was performed at an amplitude of γ = 100% for 60 s and during which G′ falls below G″, confirming the breakdown of the gel network. Resetting the strain to 0.1%, G′ of both hydrogels rapidly exceeded G″ within 10 s. The G-C18:1/CNC$_{0.2}$ hydrogel exhibited a recovery of 78% after 20 seconds and 90% after 10 min, while the G-C18:1/SCNC hydrogel showed a recovery of 70% after 20 s and 84% after 10 min. These findings demonstrate the thixotropic recovery behavior of these hydrogels, which is beneficial for applications requiring injectability or spreading.[49] The observed differences in recovery rates and mechanical strength between the G-C18:1/CNC$_{0.2}$ and G-C18:1/SCNC hydrogels are typical for soft gel systems.[50,51]

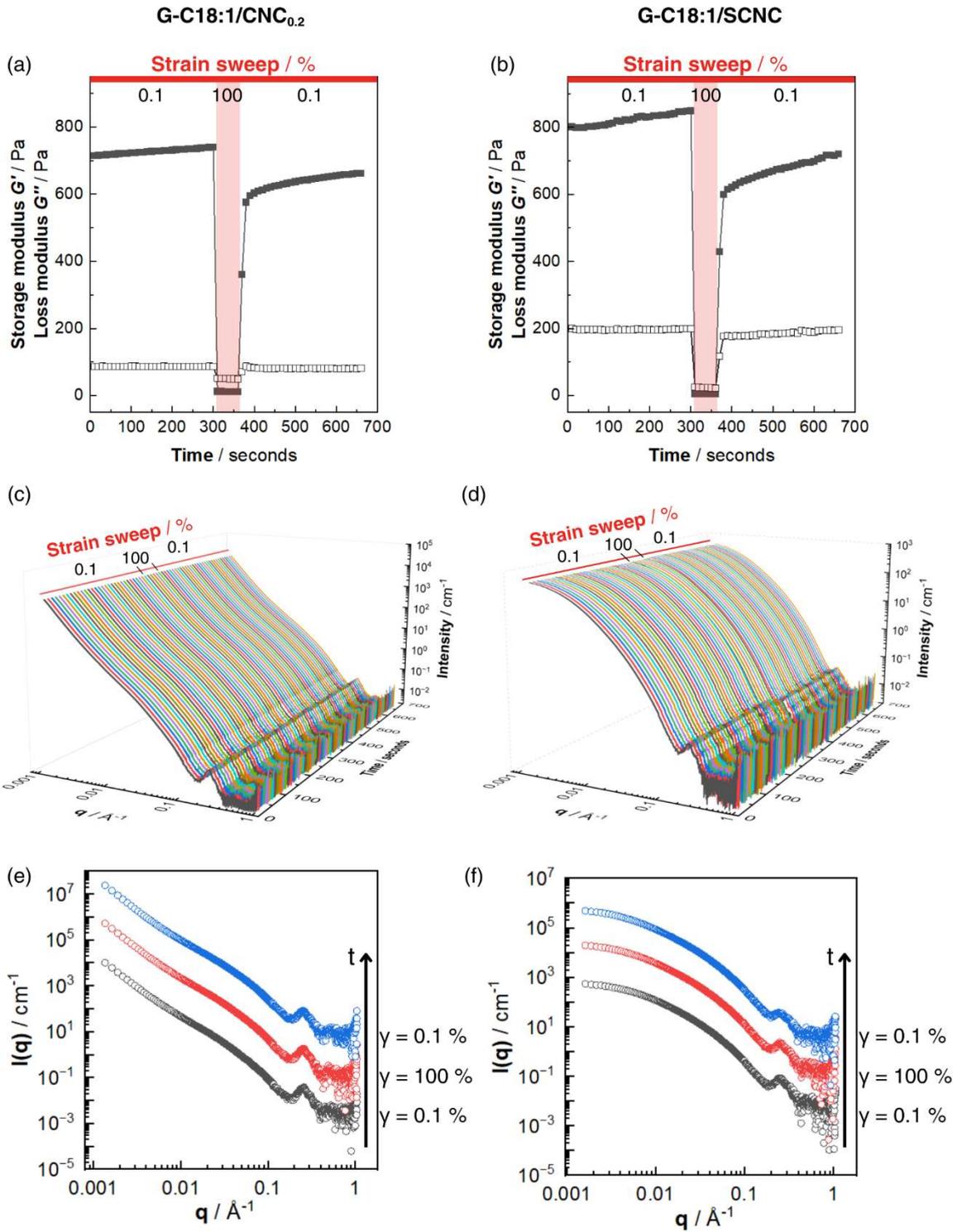

**Figure 5.** Study of the recovery properties of G-C18:1/CNC$_{0.2}$ (left) and G-C18:1/SCNC (right) hydrogels after a cycle of rupture-recovery. (a, b) Evolution of G′ (full squares) and G″ (empty red squares) in a typical strain sweep experiment performed with an initial strain of γ = 0.1% followed by a rupture for 1 minute at γ = 100% and recovery of 5 minutes at γ = 0.1%. (c, d) Full SAXS data recorded during the experiment at the SWING beamline of Soleil synchrotron in radial position. (e, f) Extracted typical SAXS profiles representing each event, shifted by a factor of 1000 for clarity.

Under rupture and during recovery phases, the SAXS profiles of two hydrogels were recorded in tangential (Figure S9) and radial positions (Figure 5c,d). As a first observation, the scattering profiles are not affected by deformation, indicating that the internal structure of the gels is resilient to strain sweep. A closer examination at specific strain values, 0.1%, 100% and eventually back to 0.1% (Figure 5e,f) shows no major changes in the corresponding SAXS profiles, which cover the length scale between 314 nm and 6 nm, and confirming that the structure of the gels is essentially unchanged and isotropic (i.e., randomly oriented) throughout the thixotropy test. An attempt to fit the data using the Ornstein–Zernike equation was made to determine the mesh size ($\xi$) of the hydrogel network, a method commonly used for swollen gels.[52] The detailed calculation is provided in Figure S10 and Table S2 of the Supporting Information for G-C18:1/SCNC hydrogel, as their SAXS profiles reach a pseudo plateau at low-q values. If $\xi$ varies from 201.2 Å ($\gamma$ = 0.1 %) to 210.6 Å ($\gamma$ = 100 %) and eventually 230.2 Å ($\gamma$ = 0.1 %) during the test cycle, one cannot reasonably correlate such variation with the thixotropic behavior at low and high strain. For this reason, the evolution of $\xi$ was not studied further.

*Orthogonal G-C18:1/SCNCs gels are temperature and pH responsive*

The advantage of orthogonal hydrogels is the contribution of the functionality of each system to the material properties.[53,54] This is shown hereafter in the context of temperature and pH dependency of the elastic properties. Previous work has shown that salt-free G-C18:1 (powder) has a main transition temperature ($T_m$) of 68.1 °C, with a pre-transition occurring at 37.6 °C.[38] While CNCs demonstrate excellent thermal stability—resisting degradation up to approximately 260 °C [37,55]—the thermal stability of SCNCs is significantly lower (<200 °C).[56] This reduction is attributed to catalytic autoxidation, driven by the acidic nature of sulfate half-ester groups on the SCNCs surface.[57] The evolution of the storage and loss moduli of G-C18:1, SCNCs, and G-C18:1/SCNC gels was examined over a heating–cooling cycle between 23 and 70 °C, as shown in Figure 6. In Figure 6a, G-C18:1 begins to lose its elastic properties around 27 °C, with complete gel breakdown occurring at 54 °C. This result aligns with previous studies on G-C18:1 gels, where temperature-dependent fiber-to-micelle phase transition drives the gel-to-sol transition (not reversible within the time frame, 40 min, explored here),[38] and with most molecular gels.[58] Figure 6b shows the essentially temperature-independent behavior of the SCNCs gel, despite a minor increase in G′ upon cooling, likely due to improved gel reorganization.[44]

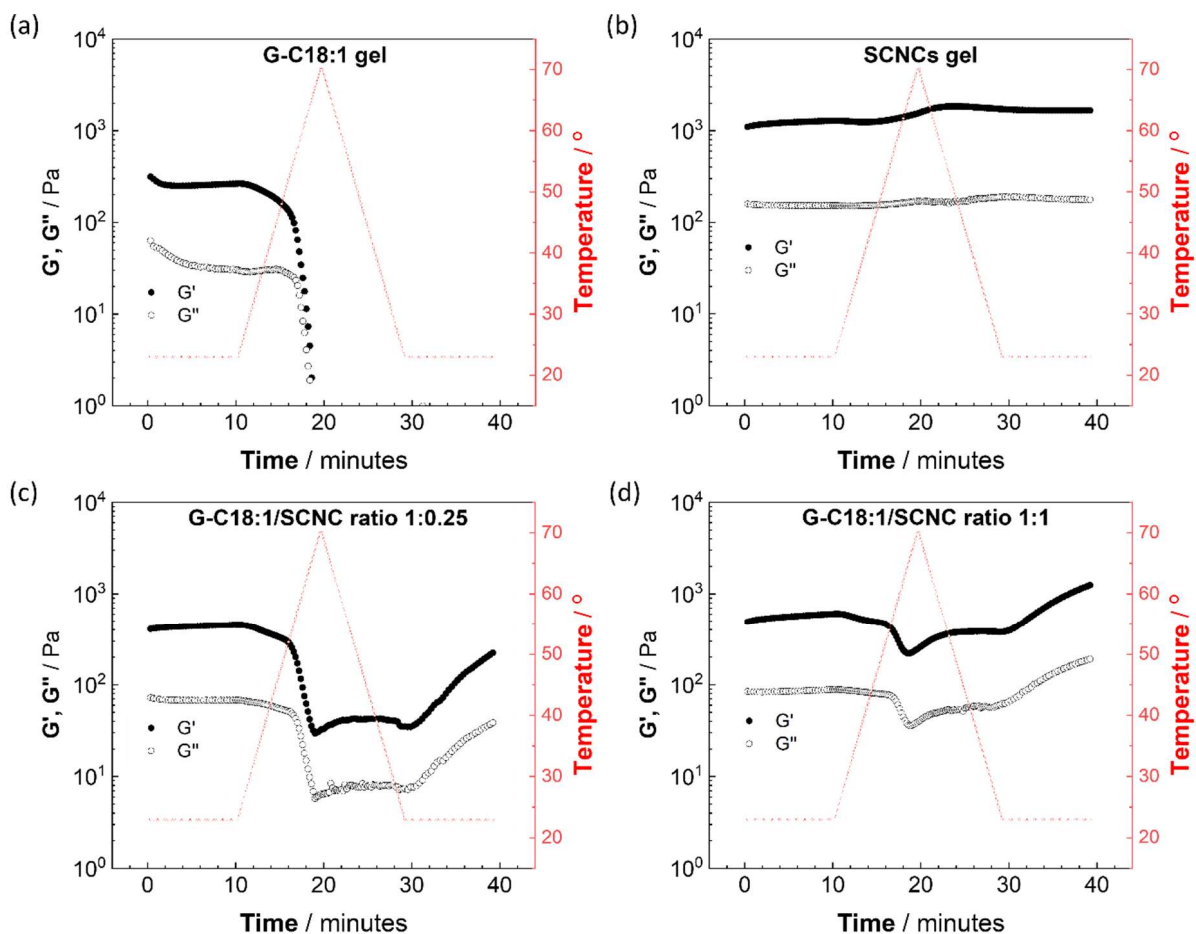

**Figure 6. Temperature responsivity of G-C18:1/SCNC gels.** Evolution of storage (G′) and loss (G″) moduli over time and temperature of (a) G-C18:1 and (b) SCNCs control hydrogels, (c,d) orthogonal G-C18:1/SCNC gels.

The temperature-dependent viscoelasticity of G-C18:1/SCNC gels largely depends on the SCNCs content (Figure 6c,d). In the 1:0.25 mass ratio hydrogel, where SCNCs primarily acts as individual reinforcements within the G-C18:1 gel matrix, the gel properties are largely dominated by the temperature-dependent behavior of the G-C18:1 network. However, it is remarkable that, unlike the G-C18:1 control gel, the dual network structure is preserved throughout the temperature test range: at 70 °C, both G′ and G″ reach their minimum values but with G′ > G″, no evolution was observed throughout the cooling cycle. The elastic properties eventually recover when the temperature returns to 23 °C, indicating a temperature-reversible behavior. A similar trend is observed in hydrogels with higher SCNC content (mass ratio 1:1), but with even less temperature sensitivity, most likely explained by the higher stability of the SCNCs network. Very interestingly, the time-dependent recovery of the gel′s elasticity continues beyond the current explored time scale, whereas for the hydrogel with a 1:1, the G′ value largely exceeds not only the initial value prior to cooling, but also the limits of the SCNCs control gels (Figure 6b). These findings demonstrate the mutual benefits of orthogonal G-C18:1/SCNC hydrogels: the SCNC network enhances the thermal stability of G-C18:1, while G-C18:1 not only modulates the viscoelasticity of SCNCs but it also improves the overall elastic properties after cooling, compared to SCNCs alone. This behavior is similar to the role of the alginate network observed in our previous study.[59]

The pH-sensitivity of these gels is another intriguing aspect to explore. pH can affect the swelling properties of CNCs gels[22,60] but it also influences the phase diagram of G-C18:1:[31] at basic pH, G-C18:1 assembles in the form of micelles, which change into fibers upon adding a $Ca^{2+}$ source.[32] Below pH 6.2, G-C18:1 assembles into vesicles[32] and below pH 4, they stabilize a lamellar precipitate.[31] Consequently, pH variations are expected to have a significant impact on the orthogonal hydrogels. Figure 7a presents the viscoelasticity of G-C18:1 with its SCNCs orthogonal gels at pH 8, then adjusted to pH 5, and finally return to pH 8. At pH 8, all samples exhibit gel-like behavior, which is lost upon decreasing the pH to 5 by adding microliter-scale amounts of 1 M HCl solution. Interestingly, although the SCNC control remains in a gel state at both pH 8 and pH 5,[61] its orthogonal network with G-C18:1 collapse, as seen in the image on Figure 7. When the pH is restored to 8, both hydrogels tested here, at low (1:0.25) and high (1:15) G-C18:1/SCNC ratio recover their solid-like properties (Figure 7b,c).

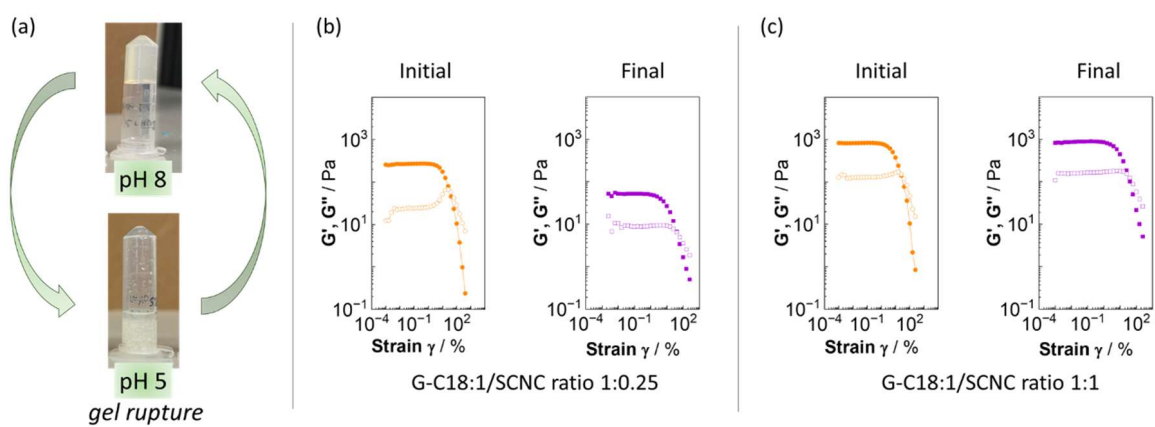

**Figure 7. pH responsivity of G-C18:1/SCNC gels. (a) Illustration of gel behavior under pH cycling from pH 8 to pH 5 and back to pH 8; (b, c) Viscoelastic behavior of G-C18:1 gels with SCNC at mass ratios of 1:0.25 (b) and 1:1 (c), showing storage modulus (G′) and loss modulus (G″) at pH 8 in the initial stage and after completing the pH cycle (final stage).**

*Perspectives of using orthogonal G-C18:1/CNCs gels as soft printable scaffolds*

Tunability of the rheo-elastic properties of the G-C18:1/CNCs orthogonal gels shown in this work and combined with their bio-based origin matches most of the requirements (degradability, injectability, tunable mechanical strength, and ease of handling) needed from hydrogels for applications in tissue engineering.[62] Benefits of the current systems are their rapid sol–gel transition (within minutes) with high yield stress, which make them shapeable and avoiding collapse under their own weight (Figure 8a). However, the fast recovery after shearing is another asset, lending injectability to the gels, regardless of whether the CNCs were produced via liquid-phase ($CNC_{0.2}$) or gas-phase hydrolysis ($gCNC_{0.2}$) (Figure 8b). As a matter of fact, in a current study it was possible to print self-standing structures by using an additive-free G-C18:1 hydrogel,[63] but at concentrations above 10 wt%. The present work shows how similar properties can be achieved by reducing the glycolipid concentration of a factor 5 and replacing with an equivalent amount of CNCs.[51,64]

These results highlight the potential of orthogonal G-C18:1/CNC gels as potential soft printable scaffolds for biomedical and pharmaceutical applications. The use of CNCs and the bio-amphiphile G-C18:1 offers a bio-based, non-toxic alternative. As reported by Peppas and coworkers,[65] most toxicity concerns related to hydrogel carriers arise from unreacted monomers, oligomers, and initiators that leach out during use.[60] Using CNCs and the bio-amphiphile G-C18:1, which are both bio-based and free of toxic monomers, enable the development of more complex bio-based systems. Moreover, the strategy of using a bio-based glycolipid as a surface stabilizer avoids the need for additional chemical modification of CNC surfaces. Chemical modification of the nanocellulose surface is necessary, but it may affect the biodegradability and potentially introduce cytotoxicity.[66] The use of a bio-based glycolipid as a surface stabilizer, as shown in previous work[34] and as repeated in this study,[77] is an interesting approach to enable the dispersion and stabilization of uncharged CNCs without the need for conventional carboxylate or sulfate surface functionalization. The other functionalities of G-C18:1, namely its gel scaffolding and rheo-modifying properties combined with various forms of CNCs, make the present orthogonal system attractive for further soft material development for potential applications like drug delivery,[64,67] or cell encapsulation.[66,68,69]

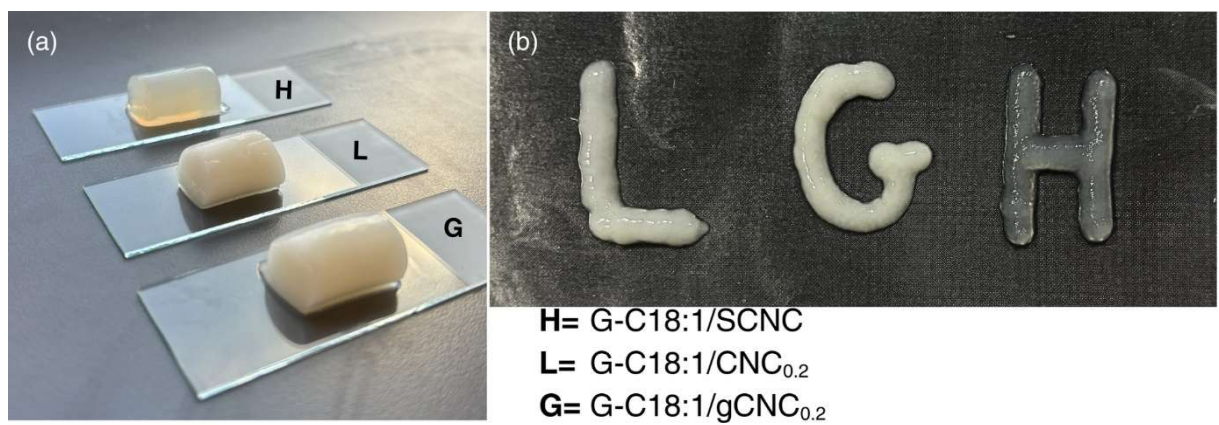

**Figure 8.** Images of G-C18:1/SCNC (H), G-C18:1/CNC$_{0.2}$ (L), and G-C18:1/gCNC$_{0.2}$ (G) hydrogels: (a) in cylindrical form and (b) injected into specific shapes using a syringe.

An additional benefit of the orthogonal nature is certainly the enhanced resistance to temperature, making such systems of potential use for cell culture applications at 37 °C and physiological pH.[70] Furthermore, these orthogonal hydrogels exhibit stimuli-responsive behavior to both temperature and pH. In a 2019 review, Fu et al. summarized stimuli-responsive cellulose-based hydrogels for biomedical applications, highlighting that CNCs are typically combined with polymers in all reported cases.[22,71] In this context, the incorporation of G-C18:1 represents a significant advancement toward the development of biosourced and more sustainable materials. Future studies could further explore the swelling behavior and assess cell viability, as well as drug delivery rate to fully evaluate the potential of this orthogonal hydrogel system in biomedical applications.

**Conclusion**

Despite growing interest in multicomponent hydrogels, the orthogonal design of colloidal and LMWG systems, particularly with bio-based, sustainable components, has received little attention. This work presents the first comprehensive study of orthogonal hydrogels formed between cellulose nanocrystals (CNCs) and a biosurfactant, exploring both matrix/reinforcement and orthogonal hydrogel systems. Beyond demonstrating that the presence of CNCs significantly enhances the mechanical properties of G-C18:1 hydrogel, this study also investigates their responsiveness to external stimuli (pH and temperature), thereby expanding the potential applications of this class of bio-based materials. SAXS data reveal an isotropic structure with a mesh size suitable for the delivery of most small-molecule drugs. The entire bio-based origin of the materials, the simplicity in terms of their synthesis and multifunctionality then further position these orthogonal gels as promising candidates as advanced bio-based soft materials. This study lays the groundwork for future research into the use of biosurfactant-nanocellulose systems in biomedical applications, opening avenues for deeper and more targeted investigations such as cell encapsulation and drug delivery. Finally, we hope this work highlights the untapped potential of microbial biosurfactant-based hydrogels in the biomedical field, helping them gain broader recognition in a market still dominated by synthetic surfactants.

**Funding Sources**

Magnus Ehrnrooth Foundation (Finland) is warmly acknowledged for funding the PhD project. The French Agence Nationale de la Recherche, project number ANR-22-CE43-0017, is acknowledged for its financial support.

**Acknowledgement**

Authors acknowledge the French Groupement de recherche (GDR) SLAMM (Solliciter LA Matière Molle) for the opportunity to run the rheo-SAXS experiment at the SWING Beamline of Synchrotron SOLEIL within the framework of the Block allocating Group (BAG), proposal No. 20231446. Dr. François Boué, CEA, Saclay, France and Dr. Thomas Gibaud, ENS, Lyon, France, are kindly acknowledge for their assistance and help during the rheo-SAXS experiment. This work is part of the FinnCERES Flagship ecosystem.

**Supporting Information**

Visual experiments for CNCs stabilization, supporting photos of gels, supplementary rheometer and SAXS data.

**Abbreviations**

CNCs, cellulose nanocrystals; gCNCs, gaseous HCl hydrolysis CNCs; HC, hydrolyzed cellulosic; LMWG, low molecular weight gelator; SCNCs, sulfuric acid hydrolysis CNCs; SAFiN, self-assembled fibrillar network; SAXS, small-angle X-ray scattering.

**References**

1. Ahmed EM. Hydrogel: Preparation, characterization, and applications: A review. *J Adv Res*. 2015;6(2):105-121. doi:10.1016/j.jare.2013.07.006

2. Chamkouri H. A Review of Hydrogels, Their Properties and Applications in Medicine. *Am J Biomed Sci Res*. 2021;11(6):485-493. doi:10.34297/AJBSR.2021.11.001682

3. Ozcelik B. Degradable hydrogel systems for biomedical applications. In: *Biosynthetic Polymers for Medical Applications*. Woodhead Publishing; 2016:173-188. doi:10.1016/B978-1-78242-105-4.00007-9

4. Feksa LR, Troian EA, Muller CD, Viegas F, Machado AB, Rech VC. Hydrogels for biomedical applications. In: *Nanostructures for the Engineering of Cells, Tissues and Organs*. Elsevier; 2018:403-438. doi:10.1016/B978-0-12-813665-2.00011-9

5. Li M, He X, Zhao R, Shi Q, Nian Y, Hu B. Hydrogels as promising carriers for the delivery of food bioactive ingredients. *Front Nutr*. 2022;9. doi:10.3389/fnut.2022.1006520

6. Nath PC, Debnath S, Sridhar K, Inbaraj BS, Nayak PK, Sharma M. A Comprehensive Review of Food Hydrogels: Principles, Formation Mechanisms, Microstructure, and Its Applications. *Gels*. 2022;9(1):1. doi:10.3390/gels9010001

7. Cornwell DJ, Smith DK. Expanding the scope of gels – combining polymers with low-molecular-weight gelators to yield modified self-assembling smart materials with high-tech applications. *Mater Horiz*. 2015;2(3):279-293. doi:10.1039/C4MH00245H

8. Morris KL, Chen L, Raeburn J, et al. Chemically programmed self-sorting of gelator networks. *Nat Commun*. 2013;4(1):1480. doi:10.1038/ncomms2499

9. Brizard A, Stuart M, van Bommel K, Friggeri A, de Jong M, van Esch J. Preparation of Nanostructures by Orthogonal Self-Assembly of Hydrogelators and Surfactants. *Angewandte Chemie International Edition*. 2008;47(11):2063-2066. doi:10.1002/anie.200704609

10. Swain JWR, Yang CY, Hartgerink JD. Orthogonal Self-Assembly of Amphiphilic Peptide Hydrogels and Liposomes Results in Composite Materials with Tunable Release Profiles. *Biomacromolecules*. 2023;24(11):5018-5026. doi:10.1021/acs.biomac.3c00664

11. Li SL, Xiao T, Lin C, Wang L. Advanced supramolecular polymers constructed by orthogonal self-assembly. *Chem Soc Rev*. 2012;41(18):5950. doi:10.1039/c2cs35099h

12. Weiner AL, Carpenter-Green SS, Soehngen EC, Lenk RP, Popescu MC. Liposome–Collagen Gel Matrix: A Novel Sustained Drug Delivery System. *J Pharm Sci*. 1985;74(9):922-925. doi:10.1002/jps.2600740903

13. Attia L, Chen L, Doyle PS. Orthogonal Gelations to Synthesize Core–Shell Hydrogels Loaded with Nanoemulsion-Templated Drug Nanoparticles for Versatile Oral Drug Delivery. *Adv Healthc Mater*. 2023;12(31). doi:10.1002/adhm.202301667

14. Draper ER, Adams DJ. Low-Molecular-Weight Gels: The State of the Art. *Chem*. 2017;3(3):390-410. doi:10.1016/j.chempr.2017.07.012

15. Yu G, Yan X, Han C, Huang F. Characterization of supramolecular gels. *Chem Soc Rev*. 2013;42(16):6697. doi:10.1039/c3cs60080g

16. Draper ER, Su H, Brasnett C, et al. Opening a Can of Worm(-like Micelle)s: The Effect of Temperature of Solutions of Functionalized Dipeptides. *Angewandte Chemie*. 2017;129(35):10603-10606. doi:10.1002/ange.201705604

17. Smith DK. Supramolecular gels – a panorama of low-molecular-weight gelators from ancient origins to next-generation technologies. *Soft Matter*. 2024;20(1):10-70. doi:10.1039/D3SM01301D

18. Brizard AM, Stuart MCA, van Esch JH. Self-assembled interpenetrating networks by orthogonal self assembly of surfactants and hydrogelators. *Faraday Discuss*. 2009;143:345. doi:10.1039/b903806j

19. Seyrig C, Poirier A, Perez J, Bizien T, Baccile N. Interpenetrated Biosurfactant–Biopolymer Orthogonal Hydrogels: The Biosurfactant's Phase Controls the Hydrogel's Mechanics. *Biomacromolecules*. 2023;24(1):33-42. doi:10.1021/acs.biomac.2c00319

20. Vieira VMP, Hay LL, Smith DK. Multi-component hybrid hydrogels – understanding the extent of orthogonal assembly and its impact on controlled release. *Chem Sci*. 2017;8(10):6981-6990. doi:10.1039/C7SC03301J

21. Onogi S, Shigemitsu H, Yoshii T, et al. In situ real-time imaging of self-sorted supramolecular nanofibres. *Nat Chem*. 2016;8(8):743-752. doi:10.1038/nchem.2526

22. Fu LH, Qi C, Ma MG, Wan P. Multifunctional cellulose-based hydrogels for biomedical applications. *J Mater Chem B*. 2019;7(10):1541-1562. doi:10.1039/C8TB02331J

23. Du H, Liu W, Zhang M, Si C, Zhang X, Li B. Cellulose nanocrystals and cellulose nanofibrils based hydrogels for biomedical applications. *Carbohydr Polym*. 2019;209:130-144. doi:10.1016/j.carbpol.2019.01.020

24. Yang J, Han CR, Duan JF, et al. Synthesis and characterization of mechanically flexible and tough cellulose nanocrystals–polyacrylamide nanocomposite hydrogels. *Cellulose*. 2013;20(1):227-237. doi:10.1007/s10570-012-9841-y

25. Dash R, Foston M, Ragauskas AJ. Improving the mechanical and thermal properties of gelatin hydrogels cross-linked by cellulose nanowhiskers. *Carbohydr Polym*. 2013;91(2):638-645. doi:10.1016/j.carbpol.2012.08.080

26. Liu YJ, Cao WT, Ma MG, Wan P. Ultrasensitive Wearable Soft Strain Sensors of Conductive, Self-healing, and Elastic Hydrogels with Synergistic "Soft and Hard" Hybrid Networks. *ACS Appl Mater Interfaces*. 2017;9(30):25559-25570. doi:10.1021/acsami.7b07639

27. Shao C, Wang M, Meng L, et al. Mussel-Inspired Cellulose Nanocomposite Tough Hydrogels with Synergistic Self-Healing, Adhesive, and Strain-Sensitive Properties. *Chem Mater*. 2018;30(9):3110-3121. doi:10.1021/acs.chemmater.8b01172

28. Poirier A, Le Griel P, Perez J, Hermida-Merino D, Pernot P, Baccile N. Metallogels from a Glycolipid Biosurfactant. *ACS Sustain Chem Eng*. 2022;10(50):16503-16515. doi:10.1021/acssuschemeng.2c01860

29. Baccile N, Poirier A, Seyrig C, et al. Chameleonic amphiphile: The unique multiple self-assembly properties of a natural glycolipid in excess of water. *J Colloid Interface Sci*. 2023;630:404-415. doi:10.1016/j.jcis.2022.07.130

30. Baccile N, Cuvier AS, Prévost S, et al. Self-Assembly Mechanism of pH-Responsive Glycolipids: Micelles, Fibers, Vesicles, and Bilayers. *Langmuir*. 2016;32(42):10881-10894. doi:10.1021/acs.langmuir.6b02337

31. Baccile N, Selmane M, Le Griel P, et al. PH-Driven Self-Assembly of Acidic Microbial Glycolipids. *Langmuir*. 2016;32(25):6343-6359. doi:10.1021/acs.langmuir.6b00488

32. Poirier A, Le Griel P, Hoffmann I, et al. Ca2+ and Ag+ orient low-molecular weight amphiphile self-assembly into "nano-fishnet" fibrillar hydrogels with unusual β-sheet-like raft domains. *Soft Matter*. 2023;19(3):378-393. doi:10.1039/D2SM01218A

33. Poirier A, Le Griel P, Perez J, Baccile N. Cation-Induced Fibrillation of Micro-bial Glycolipid Biosurfactant Probed by Ion-Resolved In Situ SAXS. *J Phys Chem B*. 2022:126. doi:10.1021/acs.jpcb.2c03739ï

34. Phi. Thuy-Linh, Xu. W., Pernot. P., Baccile N., Kontturi E. Enhancing aqueous dispersibility of uncharged cellulose through biosurfactant adsorption. *ACS Appl Polym Mater*. Published online 2025.

35. Rusli R, Shanmuganathan K, Rowan SJ, Weder C, Eichhorn SJ. Stress Transfer in Cellulose Nanowhisker Composites—Influence of Whisker Aspect Ratio and Surface Charge. *Biomacromolecules*. 2011;12(4):1363-1369. doi:10.1021/bm200141x

36. Klemm D, Kramer F, Moritz S, et al. Nanocelluloses: A new family of nature-based materials. *Angew Chem Int Ed*. 2011;50(24):5438-5466. doi:10.1002/anie.201001273

37. Thomas B, Raj MC, Athira BK, et al. Nanocellulose, a Versatile Green Platform: From Biosources to Materials and Their Applications. *Chem Rev*. 2018;(24):11575-11625. doi:10.1021/acs.chemrev.7b00627

38. Poirier A, Le Griel P, Bizien T, Zinn T, Pernot P, Baccile N. Shear recovery and temperature stability of Ca2+ and Ag+ glycolipid fibrillar metallogels with unusual β-sheet-like domains. *Soft Matter*. 2023;19(3):366-377. doi:10.1039/D2SM00374K

39. Marcovich NE, Auad ML, Bellesi NE, Nutt SR, Aranguren MI. Cellulose micro/nanocrystals reinforced polyurethane. *J Mater Res*. 2006;21(4):870-881. doi:10.1557/jmr.2006.0105

40. Pooyan P, Kim IT, Jacob KI, Tannenbaum R, Garmestani H. Design of a cellulose-based nanocomposite as a potential polymeric scaffold in tissue engineering. *Polymer (Guildf)*. 2013;54(8):2105-2114. doi:10.1016/j.polymer.2013.01.030

41. Liu H, Li C, Wang B, et al. Self-healing and injectable polysaccharide hydrogels with tunable mechanical properties. *Cellulose*. 2018;25(1):559-571. doi:10.1007/s10570-017-1546-9

42. Pääkkönen T, Spiliopoulos P, Knuts A, et al. From vapour to gas: optimising cellulose degradation with gaseous HCl. *React Chem Eng*. 2018;3(3):312-318. doi:10.1039/C7RE00215G

43. De France KJ, Hoare T, Cranston ED. Review of Hydrogels and Aerogels Containing Nanocellulose. *Chem Mater*. 2017;29(11):4609-4631. doi:10.1021/acs.chemmater.7b00531

44. Lu S, Hu X, Xu B, et al. Effects of different salt ions on the structure and rheological behavior of sulfated cellulose nanocrystal hydrogel. *Food Hydrocoll*. 2024;151:109799. doi:10.1016/j.foodhyd.2024.109799

45. Chau M, Sriskandha SE, Pichugin D, et al. Ion-Mediated Gelation of Aqueous Suspensions of Cellulose Nanocrystals. *Biomacromolecules*. 2015;16(8):2455-2462. doi:10.1021/acs.biomac.5b00701

46. Marshall LJ, Matsarskaia O, Schweins R, Adams DJ. Enhancement of the mechanical properties of lysine-containing peptide-based supramolecular hydrogels by chemical cross-linking. *Soft Matter*. 2021;17(37):8459-8464. doi:10.1039/D1SM01136G

47. Panja S, Adams DJ. Chemical crosslinking in 'reactive' multicomponent gels. *Chemical Communications*. 2022;58(37):5622-5625. doi:10.1039/D2CC00919F

48. Laun M, Auhl D, Brummer R, et al. Guidelines for checking performance and verifying accuracy of rotational rheometers: viscosity measurements in steady and oscillatory shear (IUPAC Technical Report). *Pure Appl Chem*. 2014;86(12):1945-1968. doi:10.1515/pac-2013-0601

49. Liu M, Zeng X, Ma C, et al. Injectable hydrogels for cartilage and bone tissue engineering. *Bone Res*. 2017;5(1):17014. doi:10.1038/boneres.2017.14

50. Ospennikov AS, Chesnokov YM, Shibaev A V., Lokshin B V., Philippova OE. Nanostructured Hydrogels of Carboxylated Cellulose Nanocrystals Crosslinked by Calcium Ions. *Gels*. 2024;10(12):777. doi:10.3390/gels10120777

51. Bassan R, Varshney M, Roy S. An Amino Acid-Based Thixotropic Hydrogel: Tuning of Gel Recovery Time by Mechanical Shaking. *ChemistrySelect*. 2023;8(5). doi:10.1002/slct.202203317

52. Shibayama M. Small Angle Neutron Scattering on Gels. In: *Soft Matter Characterization*. Springer Netherlands; 2008:783-832. doi:10.1007/978-1-4020-4465-6_14

53. Morrison TX, Gramlich WM. Tunable, thiol-ene, interpenetrating network hydrogels of norbornene-modified carboxymethyl cellulose and cellulose nanofibrils. *Carbohydr Polym*. 2023;319:121173. doi:10.1016/j.carbpol.2023.121173

54. Seth P, Friedrichs J, Limasale YDP, et al. Interpenetrating Polymer Network Hydrogels with Tunable Viscoelasticity and Proteolytic Cleavability to Direct Stem Cells In Vitro. *Adv Healthc Mater*. 2025;14(9). doi:10.1002/adhm.202402656

55. Moon RJ, Martini A, Nairn J, Simonsen J, Youngblood J. Cellulose nanomaterials review: structure, properties and nanocomposites. *Chem Soc Rev*. 2011;40(7):3941. doi:10.1039/c0cs00108b

56. D'Acierno F, Hamad WY, Michal CA, MacLachlan MJ. Thermal Degradation of Cellulose Filaments and Nanocrystals. *Biomacromolecules*. 2020;21(8):3374-3386. doi:10.1021/acs.biomac.0c00805

57. Martínez-Sanz M, Lopez-Rubio A, Lagaron JM. Optimization of the nanofabrication by acid hydrolysis of bacterial cellulose nanowhiskers. *Carbohydr Polym*. 2011;85(1):228-236. doi:10.1016/j.carbpol.2011.02.021

58. Raghavan SR, Douglas JF. The conundrum of gel formation by molecular nanofibers, wormlike micelles, and filamentous proteins: gelation without cross-links? *Soft Matter*. 2012;8(33):8539. doi:10.1039/c2sm25107h

59. Seyrig C, Poirier A, Bizien T, Baccile N. In Situ Stimulation of Self-Assembly Tunes the Elastic Properties of Interpenetrated Biosurfactant–Biopolymer Hydrogels. *Biomacromolecules*. 2023;24(1):19-32. doi:10.1021/acs.biomac.2c01062

60. Ooi SY, Ahmad I, Amin MohdCIM. Cellulose nanocrystals extracted from rice husks as a reinforcing material in gelatin hydrogels for use in controlled drug delivery systems. *Ind Crops Prod*. 2016;93:227-234. doi:10.1016/j.indcrop.2015.11.082

61. Lombardo S, Gençer A, Schütz C, Van Rie J, Eyley S, Thielemans W. Thermodynamic Study of Ion-Driven Aggregation of Cellulose Nanocrystals. *Biomacromolecules*. 2019;20(8):3181-3190. doi:10.1021/acs.biomac.9b00755

62. Hoffman AS. Hydrogels for biomedical applications. *Adv Drug Deliv Rev*. 2002;54(1):3-12. doi:10.1016/S0169-409X(01)00239-3

63. Fragal EH, Poirier A, Bleses D, Faria Guimarães Silva Y, Baccile N, Rharbi Y. Microbial biosurfactant hydrogels with tunable rheology for precision 3D printing of soft scaffolds. *Soft Matter*. 2025;21(22):4476-4487. doi:10.1039/D5SM00248F

64. Yang Z, Xu K, Wang L, et al. Self-assembly of small molecules affords multifunctional supramolecular hydrogels for topically treating simulated uranium wounds. *Chem Commun*. 2005;(35):4414. doi:10.1039/b507314f

65. Peppas N. Hydrogels in pharmaceutical formulations. *Eur J Pharm Biopharm*. 2000;50(1):27-46. doi:10.1016/S0939-6411(00)00090-4

66. Seabra AB, Bernardes JS, Fávaro WJ, Paula AJ, Durán N. Cellulose nanocrystals as carriers in medicine and their toxicities: A review. *Carbohydr Polym*. 2018;181:514-527. doi:10.1016/j.carbpol.2017.12.014

67. Adetunji AI, Olaniran AO. Production and potential biotechnological applications of microbial surfactants: An overview. *Saudi J Biol Sci*. 2021;28(1):669-679. doi:10.1016/j.sjbs.2020.10.058

68. Ikeda M, Ueno S, Matsumoto S, et al. Three-Dimensional Encapsulation of Live Cells by Using a Hybrid Matrix of Nanoparticles in a Supramolecular Hydrogel. *Chem Eur J*. 2008;14(34):10808-10815. doi:10.1002/chem.200801144

69. Domingues RMA, Gomes ME, Reis RL. The Potential of Cellulose Nanocrystals in Tissue Engineering Strategies. *Biomacromolecules*. 2014;15(7):2327-2346. doi:10.1021/bm500524s

70. Wang W, Wang H, Ren C, et al. A saccharide-based supramolecular hydrogel for cell culture. *Carbohydr Res*. 2011;346(8):1013-1017. doi:10.1016/j.carres.2011.03.031

71. Hynninen V, Hietala S, McKee JR, et al. Inverse Thermoreversible Mechanical Stiffening and Birefringence in a Methylcellulose/Cellulose Nanocrystal Hydrogel. *Biomacromolecules*. 2018;19(7):2795-2804. doi:10.1021/acs.biomac.8b00392

# Supporting Information

**Orthogonality between cellulose nanocrystals and a low-molecular weight gelator**

*Thuy-Linh Phi [a,b], Eero Kontturi [b,*], Niki Baccile [a,*]*

[a] Sorbonne Université, Centre National de la Recherche Scientifique, Laboratoire de Chimie de la Matière Condensée de Paris, LCMCP, F-75005 Paris, France

[b] Department of Bioproducts and Biosystems, School of Chemical Engineering, Aalto University, P.O. Box 16300, 00076 Aalto, Finland

**Corresponding Author**
E-mail: niki.baccile@sorbonne-universite.fr
E-mail: eero.kontturi@aalto.fi

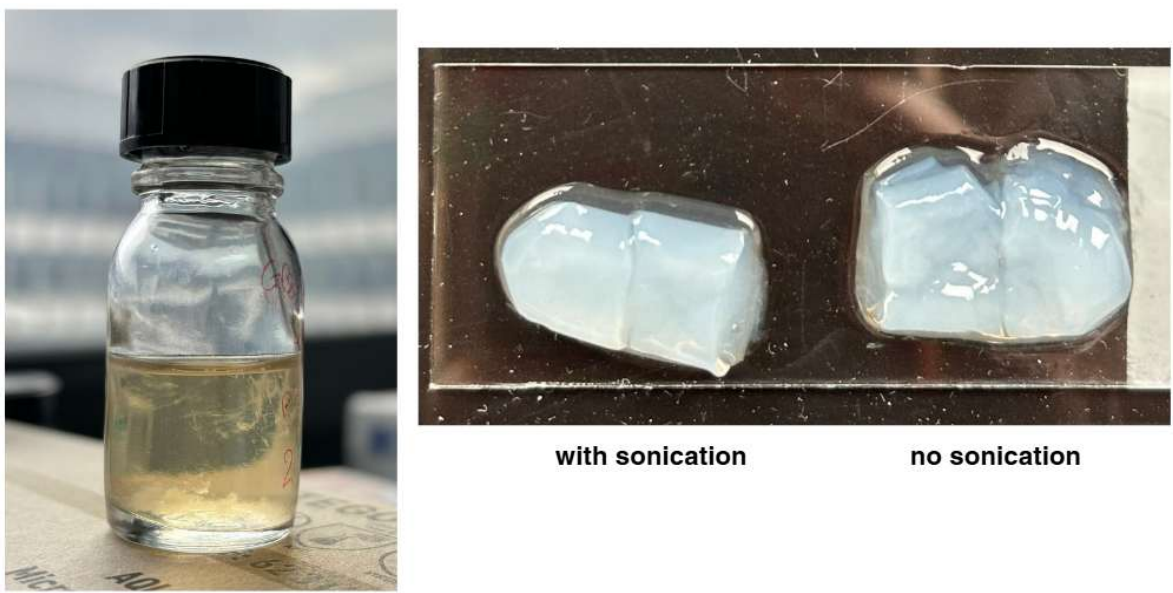

**Figure S1.** The G-C18:1 solution is prone to aggregation over time, leading to inconsistencies in gel properties. Sonication prior to each use effectively disperses aggregates, ensuring consistent gel quality.

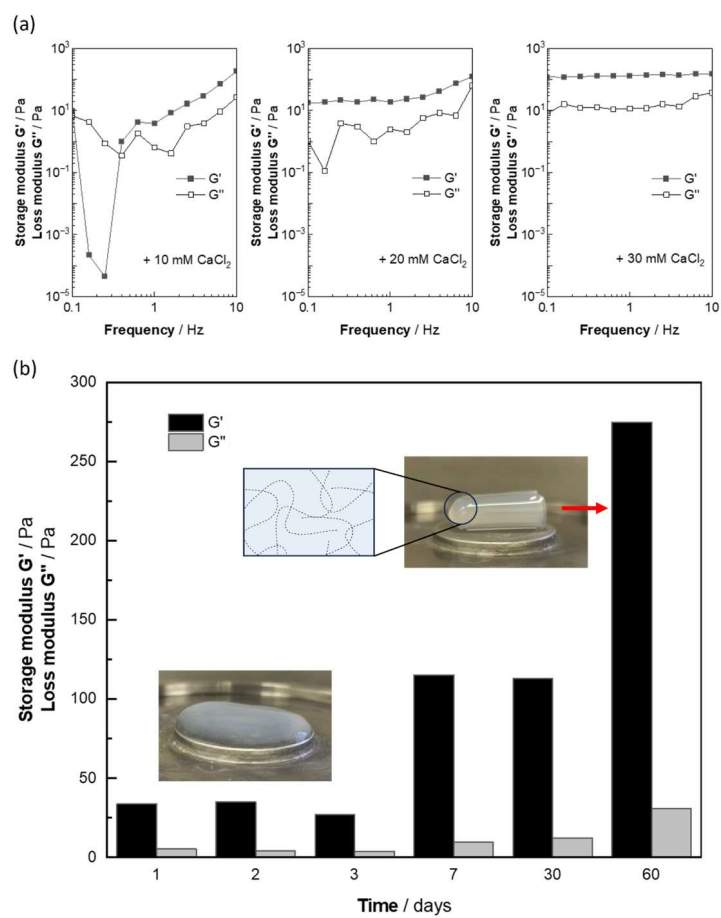

**Figure S2.** Rheological study of G-C18:1 hydrogel. (a) Typical frequency-sweep storage (full symbols) and loss (empty symbols) moduli for 2 wt% G-C18:1 with the addition of different concentrations of CaCl₂ solution (10-30 mM); and (b) time-dependency of the storage and loss moduli (f= 1 Hz, γ= 0.1 %) of G-C18:1 gels triggered by 27 mM CaCl₂ tracked for a period of 2 months.

The hydrogel properties of G-C18:1 are reproduced in this work at a fixed concentration of 2 wt% G-C18:1, close to the lower limit of 1 wt% required for hydrogel formation, while yielding the weakest mechanical behavior of the fibrillar gels.[1–3] Presumably, $Ca^{2+}$ is coordinated by the carboxylate anions of G-C18:1 at pH above 8, as the pKa of this molecule is 5.7.[2,4] The impact of the cation type, content, G-C18:1 concentration and temperature were reported in previous studies and broadly reveal that hydrogels can be formed immediately for an optimal cation-to-surfactant molar ratio, $\frac{n_{Ca^{2+}}}{n_{G-C18:1}}$, starting at 0.5, which corresponds to a positive-to-negative charge ratio of about 1.[2,5] The CaCl₂ content was then varied between 10 and 30 mM, with an approximate cation-to-surfactant molar ratio of 0.5 corresponding to ~24 mM. G-C18:1 gels triggered by 27 mM CaCl₂ were monitored over 60 days.

In agreement with previous work,[2] Figure S2a reveals low and noisy storage (G') and loss (G") moduli against frequency for a $CaCl_2$ content below the stoichiometric ratio (< 20 mM).[31] The addition of 30 mM $CaCl_2$, on the other hand, promotes stronger gelling,[6,7] of which the rheological signature displays a characteristic plateau in G' in the entire frequency range and exhibiting a G'/G" ratio greater than 10. For the rest of this work, although a minimum of 24 mM $CaCl_2$ is calculated to be enough for gelling, a slightly higher amount of 27 mM was employed to ensure consistent gelation. The mechanical properties of the G-C18:1 gel were eventually monitored over a two-month period (Figure S2b). During the first three days, G' remained relatively stable at approximately 35 Pa. However, after seven days, G' increased to 115 Pa, aligning with the storage modulus observed for the structurally similar glycolipid biosurfactant SL-C18:0 (G' ≈ 100 Pa).[8] After two months, the storage modulus of the G-C18:1 hydrogel reached 275 Pa. Considering that the evolution in time of the gel elastic properties, we adopted a compromise between gel strength and waiting time, and we decided to consistently employ a 7-days rest protocol before mixing G-C18:1 gels with the different types of CNCs. This time-dependent variation in G′ is a common characteristic of hydrogels and has been previously reported, for example, in SAFiN hydrogels of sBola C16:0 SS and gelatin gels.[5,9]

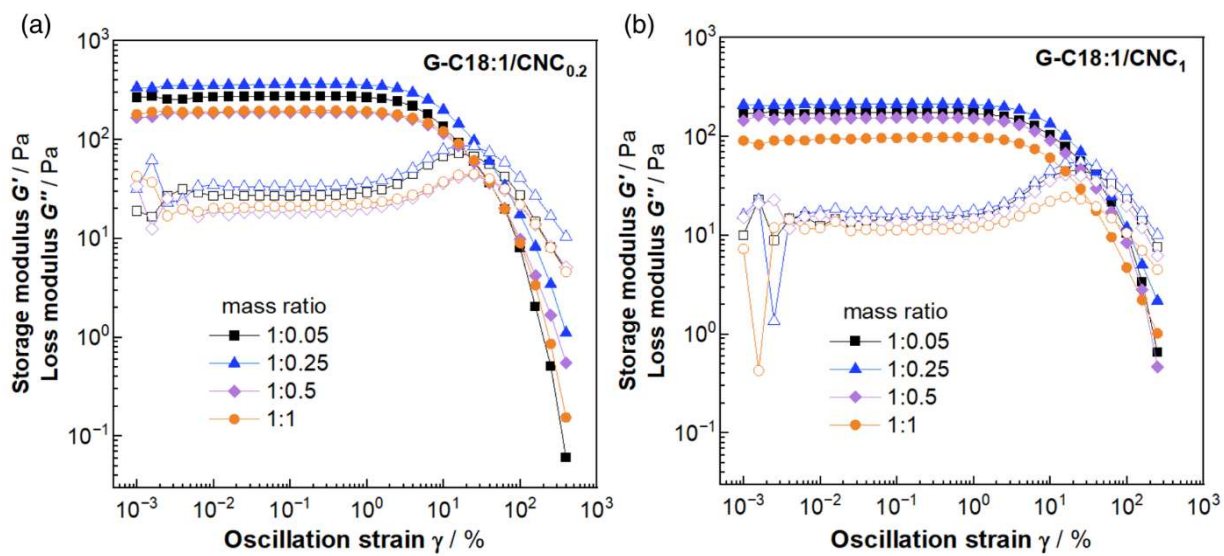

**Figure S3.** Oscillatory strain measurements at 1 Hz show the storage $G'$ (full symbols) and loss $G''$ (empty symbols) moduli for G-C18:1/CNC$_\alpha$ gels ($\alpha$= 0.2 or 1). The detailed composition for the mass ratio are given in Table 1 in the main text.

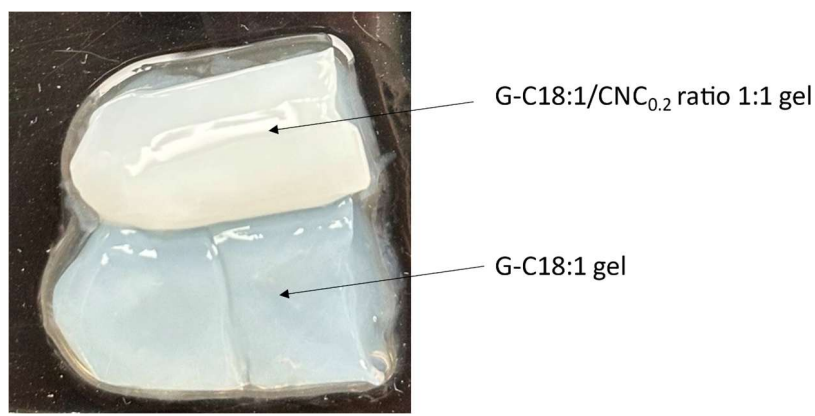

**Figure S4.** Image of G-C18:1 hydrogel and G-C18:1/CNC$_{0.2}$ 1:1 ratio gel. A clear difference in stiffness is evident, as the G-C18:1/CNC$_{0.2}$ gel exhibits a significantly firmer structure. In addition, the incorporation of CNCs into the G-C18:1 hydrogel matrix affects the transparency of the gel, leading to a milky-white appearance attributable to the CNCs component.

**Small-Angle X-ray Scattering**. Small-Angle X-ray Scattering (SAXS) measurements were performed using a Xenocs Xeuss 1.0 laboratory beamline. The instrument employed a Cu source with a wavelength of 1.54 Å and a detector distance of 310 mm and 2500 mm. High-resolution measurements were obtained with an exposure time of 1200 seconds. All SAXS experiments are performed with 1.5 mm quartz capillaries. Samples are manually injected into the capillary using a 1.0 mL syringe using a home-made device. Absolute intensity units were calibrated by subtracting the background scattering signal of water. SAXS data were analyzed using the model-independent Guinier function and the parallelepiped form factor model available in SasView freeware (version 6.0.0). For the parallelepiped model, we optimized the following parameters[10] for the commercial sulfate-rich CNCs (SCNCs): Scale: 0.001 (fixed); Background: 0.0002 cm$^{-1}$ (fixed); solvent scattering length density (sld_solvent): $9.4 \times 10^{-6}$ Å$^{-2}$ (fixed, typical for water); parallelepiped scattering length density (sld): $14.5 \times 10^{-6}$ Å$^{-2}$ (variable); length_a (height, h in main text): 20 Å (variable); length_b (width, w in main text): 410 Å (variable); length_c: 4000 Å (fixed).

In this model, the scale represents the volume fraction (0.1 w%). The parameter length_c is arbitrarily set to define an infinite length. The cellulose scattering length density (sld) was fixed at $14.5 \times 10^{-6}$ Å$^{-2}$, following the values reported by Grachev et al.[11] The SAXS profiles for SCNCs and CNCs were fitted using the same parameter set, with length_a and length_b as the only free variables.

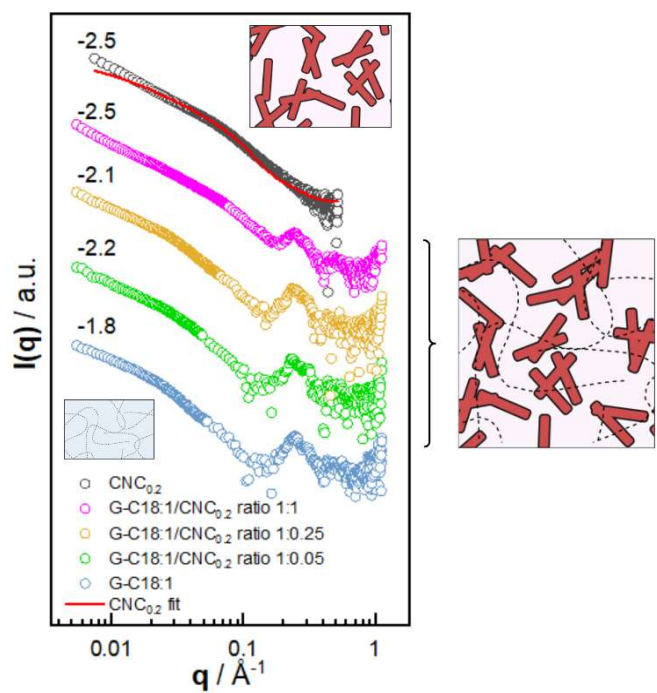

**Figure S5. Structural study of G-C18:1 hydrogel reinforced with CNC colloidal aggregates, $CNC_{0.2}$. SAXS profiles of G-C18:1, G-C18:1/$CNC_{0.2}$ at ratios of 1:0.05, 1:0.25, and 1:1, and $CNC_{0.2}$ (no G-C18:1 added) control (from bottom to top). Each curve was scaled by a factor of 1000 for clarity.**

The *log(I)–log(q)* dependency at low scattering vectors provides critical information about the physical morphology of samples, identifying specific shapes (e.g., -1 for rods, -2 for planes) or mass and surface fractals (e.g., -1 to -3 for mass, -3 to -4 for surfaces).[12]

The SAXS signal (blue profile) of G-C18:1 is in agreement with previously published data.[2] The low-q slope of approximately -1.8 was explained with the scattering signal of flat fibers,[1,3] expected to be -2. The slight discrepancy between the theoretical (-2) and experimental exponent (-1.8) values can be attributed to fiber aggregation within the hydrogel, as observed for similar systems.[5] Furthermore, the SAXS profile of G-C18:1 hydrogels lacks a hump in the mid-q region (around 0.08 Å$^{-1}$) and the oscillations above 0.1 Å$^{-1}$, typical for micellar systems. This indicates that all G-C18:1 molecules underwent the expected micelle-to-fiber transition upon addition of calcium ions.[13] Finally, the well-defined structural peak at high-q associated to the internal fiber structure and triggered by $Ca^{2+}$ is also consistent with findings from previous studies.[1,13] The SAXS profile (black curve) for $CNC_{0.2}$ shown in Figure S5 was extracted from our previous study, where this characterization has been well-described.[10] The SAXS profile of $CNC_{0.2}$ can be fitted with a classical parallelepiped form factor model[11] with values of the parallelepiped height, *h*= 35 Å, and width, *w*= 190 Å. However, the fit is poor below 0.1 nm$^{-1}$, where the slope of -2.5 indicates of large fractal interfaces,[14] which can be explained by local aggregations of CNCs.[10] To note that the SAXS profile of $CNC_{0.2}$ lacks signals of free G-C18:1, employed in this control sample as surface stabilizer.[10]

When $CNC_{0.2}$ is incorporated into the G-C18:1 hydrogels, the SAXS profiles are essentially not modified above 0.02 Å$^{-1}$ in comparison to the G-C18:1 fiber signal. This shows that the presence of CNCs neither affects the fibers morphology nor their order below the length scale of about 60 nm (q ~0.01 Å$^{-1}$). Only the SAXS profile related to the sample at 1:1 mass fraction displays a loss in the intensity of the diffraction peak and a shift in the minimum at about 0.12 Å$^{-1}$. This behavior can easily be explained by a mere effect due to the additive contribution of the individual G-C18:1 and CNCs SAXS signals, as specifically demonstrated later in the manuscript for the sulfated CNCs, here labelled SCNCs (refer to experimental section). On the other hand, the low-q slope gradually evolves from −1.8 (pure G-C18:1) to −2.5 (mass ratio sample of 1:1), this value being the same as the CNCs control. This behavior is expected and in good agreement with the increasing content of CNCs, which dominates the low-q portion of the SAXS profile. The SAXS study combined with the rheological data presents evidence that the moderate CNCs incorporation effectively reinforces and preserves the G-C18:1 network in the 1-100 nm length scale. On the other hand, the loss in mechanical strength at high CNCs concentrations (1:1 mass ratio) correlates well with the more and more dominant CNCs signal, eventually demonstrating that the elastic properties are controlled at length scales above 100 nm$^{-1}$.

S9

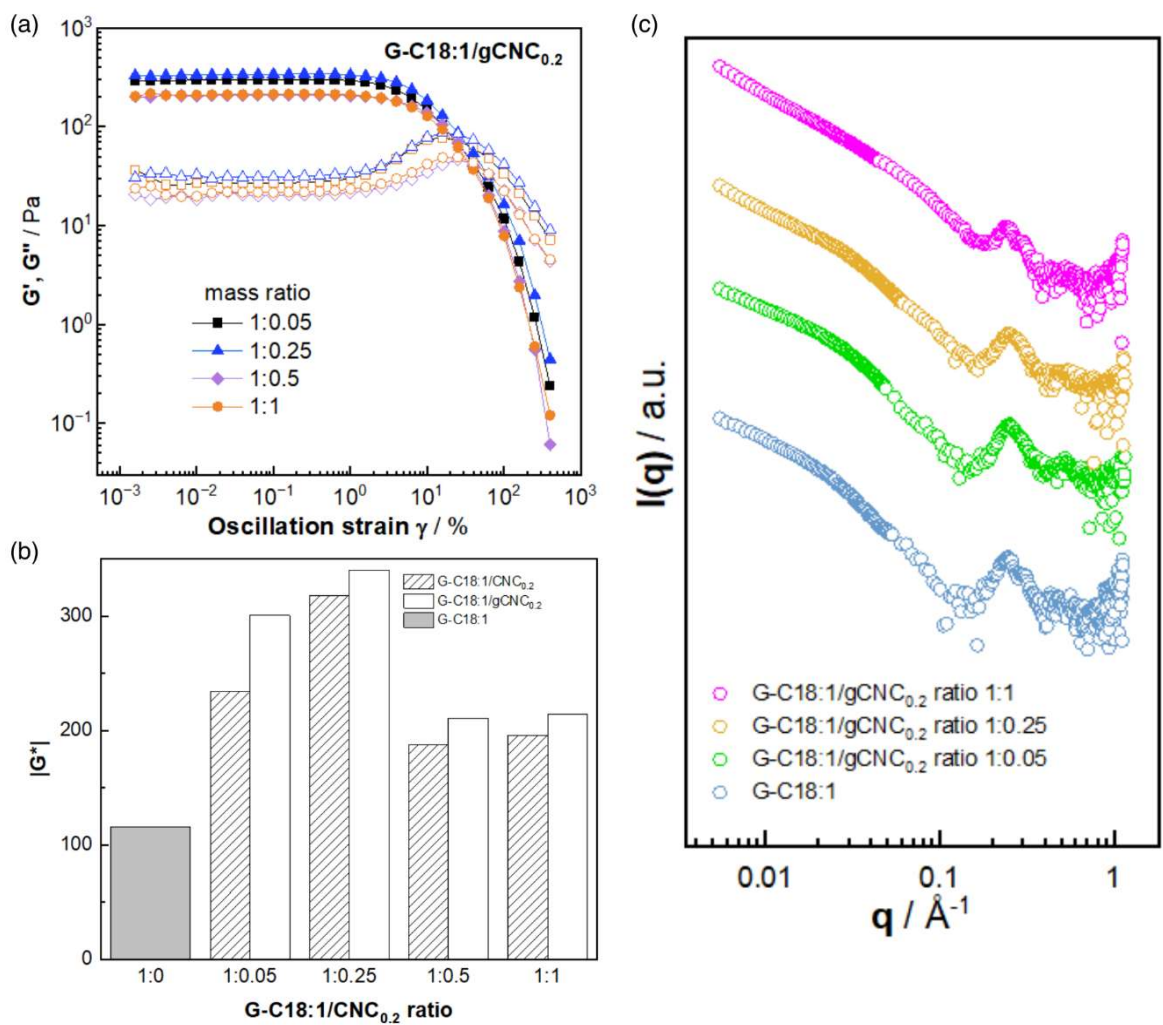

**Figure S6. G-C18:1/gCNC$_{0.2}$ gels, where CNC is prepared using gaseous hydrolysis (gCNC). (a) Oscillatory strain measurements at 1 Hz show the storage G' (full symbols) and loss G'' (empty symbols) moduli for G-C18:1/gCNC$_{0.2}$ at different ratios; (b) The complex shear modulus |G*| of G-C18:1/gCNC$_{0.2}$ and G-C18:1/gCNC$_{0.2}$ gels at different ratio for comparison, and (c) SAXS profiles of G-C18:1, G-C18:1/gCNC$_{0.2}$ at ratios of 1:0.05, 1:0.25, and 1:1, and gCNCs (from bottom to top). Each curve was scaled by a factor of 1000 for clarity.**

Hydrolysis with HCl gas to yield CNCs (gCNCs) was carried out in a custom-built reactor, following the method of Kontturi *et al.*[15] The oscillatory strain measurements for G-C18:1/gCNC$_{0.2}$ gels, shown in Figure S6a, closely resemble those of G-C18:1/CNC$_{0.2}$ (Figure S3a) and it shows the typical profile reported for LMWG: the linear viscoelastic regime ($\gamma$<10%) is characterized by G' > G'', indicating predominantly elastic behavior, and followed by an eventual loss of the elastic properties beyond $\gamma$> 100%. The key rheological parameters for G-C18:1/gCNC$_{0.2}$ gels at varying mass ratios (1:0.05, 1:0.25, 1:0.5, and 1:1) include tan δ, |G*| and critical strain ($\gamma_c$), shown in Table S1. The loss factor values remain low across all mass ratios (0.09–0.11), indicating that elastic behavior dominates over viscous behavior for

S10

all samples, as observed in G-C18:1/CNC$_{0.2}$ systems. The |G*| values lay between 210 Pa and 340 Pa, with an optimum at 1:0.25 mass ratio. Although the |G*| values of G-C18:1/gCNC$_{0.2}$ hydrogels are slightly higher than those found using aqueous CNCs as filler (Table S1 and Figure S6b). This behavior confirms that an optimal G-C18:1/gCNCs mass ratio exists so to maximize the gel's strength. The critical strain ($\gamma_c$), determined by the intersection of storage and loss modulus, shows the transition from primarily elastic to viscoelastic or flow-dominant behavior.[16] A high critical strain suggests that the material has good flexibility and maintains its structure over a widereven aft strain range.[16] In the present system, $\gamma_c$ rises as the gCNCs concentration increases, reaching a maximum (37.2 %) at a 50% gCNCs content. Interestingly, at a G-C18:1/gCNC$_{0.2}$ ratio of 1:0.5, |G*| decreases even as $\gamma_c$ reaches its maximum. The SAXS data shown in Figure S6c is similar with Figure S5 of G-C18:1/CNC$_{0.2}$, suggesting that the hydrolysis method does not significantly influence the final gel structure.

Table S1. Rheological properties of G-C18:1/gCNC$_{0.2}$ hydrogels at different mass ratios

| G-C18:1/gCNC$_{0.2}$ | ratio | | | |
|---|---|---|---|---|
| | 1:0.05 | 1:0.25 | 1:0.5 | 1:1 |
| tan δ | 0.09 | 0.09 | 0.10 | 0.11 |
| |G*| | 300.68 | 340.30 | 210.58 | 214.31 |
| $\gamma_c$ (%) | 22.9 | 26.3 | 37.2 | 32.8 |

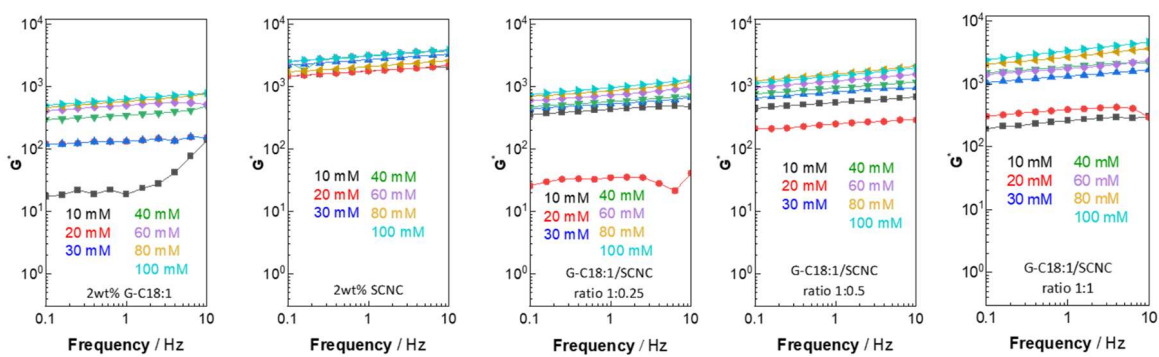

**Figure S7.** The complex shear modulus (G*) in function of frequency at varying $CaCl_2$ concentrations (10 to 100 mM), mixed with 2 wt% G-C18:1, 2 wt% SCNCs, and G-C18:1/SCNC systems. In the G-C18:1/SCNC systems, the concentration of G-C18:1 is fixed at 2 wt%, while the SCNCs concentration varies according to the mass ratio between G-C18:1 and SCNCs.

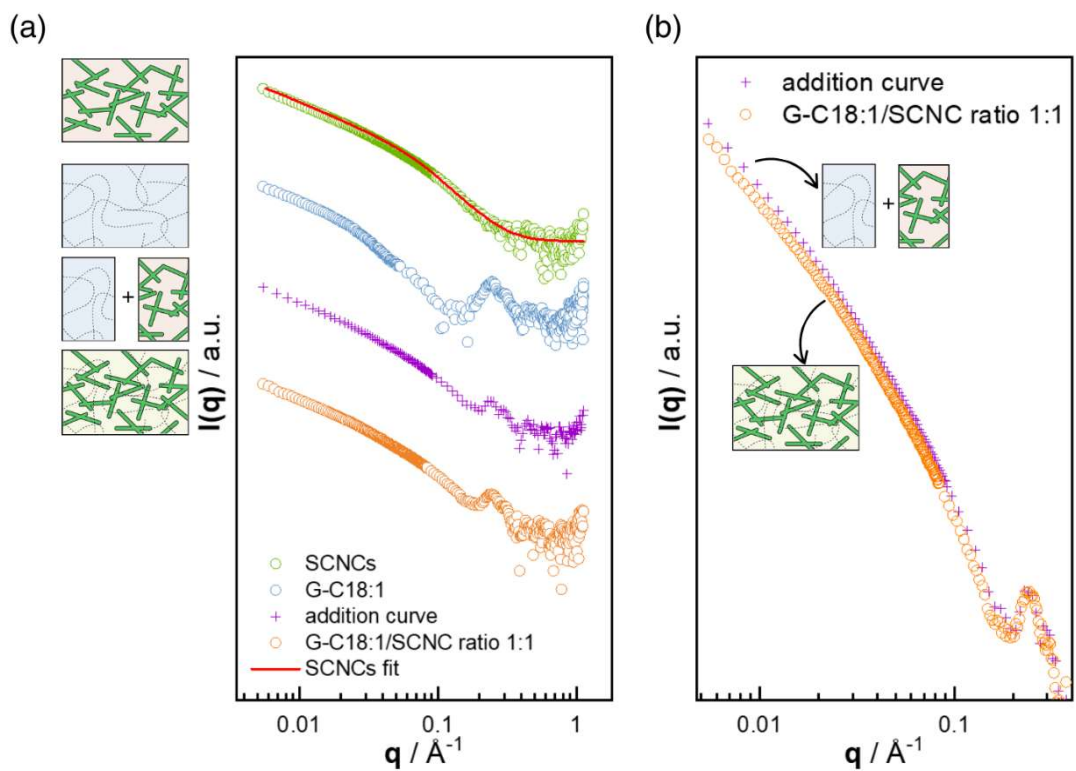

**Figure S8.** Demonstrating overlapping of G-C18:1 and SCNC networks using SAXS. (a) SAXS curves of SCNCs, G-C18:1, the arithmetic addition curve of these two components and experimental curve of the gel, where each curve was scaled by a factor of 1000 for clarity; and (b) overlap zoom-in of arithmetic addition and experimental curves at low-q.

SAXS is employed again in Figure S8 to confirm the overlapping 3D structure of G-C18:1 and SCNCs networks. The SAXS profile of SCNCs (green curve, Figure S8a) can be satisfactorily fitted with a classical parallelepiped form factor model ($h$= 21 Å, $w$= 412 Å),[11] as described for the SCNCs. The arithmetic addition of the individual SCNCs and G-C18:1 SAXS profiles (green and blue curves) results in the purple curve (Figure S8a), representing the expected SAXS intensity for a structurally overlapped G-C18:1/SCNC gel. The comparison of the added profile with the experimental G-C18:1/SCNC (1:1) data in Figure S8b shows a nearly perfect match in the entire q-range, confirming the presence of overlapped, co-existed three-dimensional networks, in agreement with previous data of G-C18:1 hydrogels with other biomacromolecules.[17]

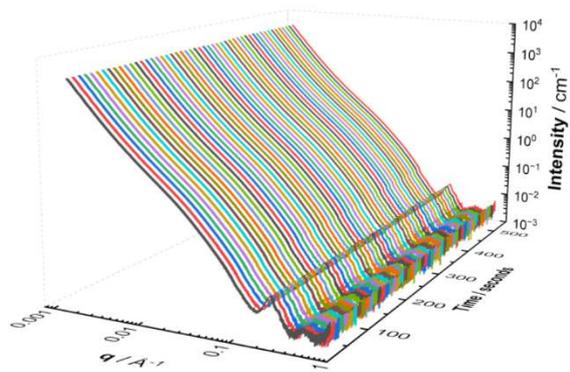 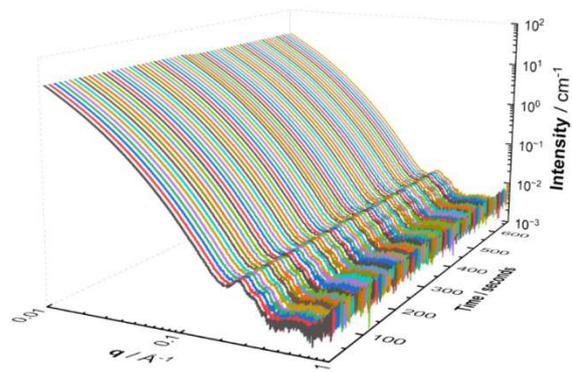

**Figure S9.** SAXS data of G-C18:1/CNC$_{0.2}$ (left) and G-C18:1/SCNC (right) recorded during the experiment at the SWING beamline of Soleil synchrotron in tangential position.

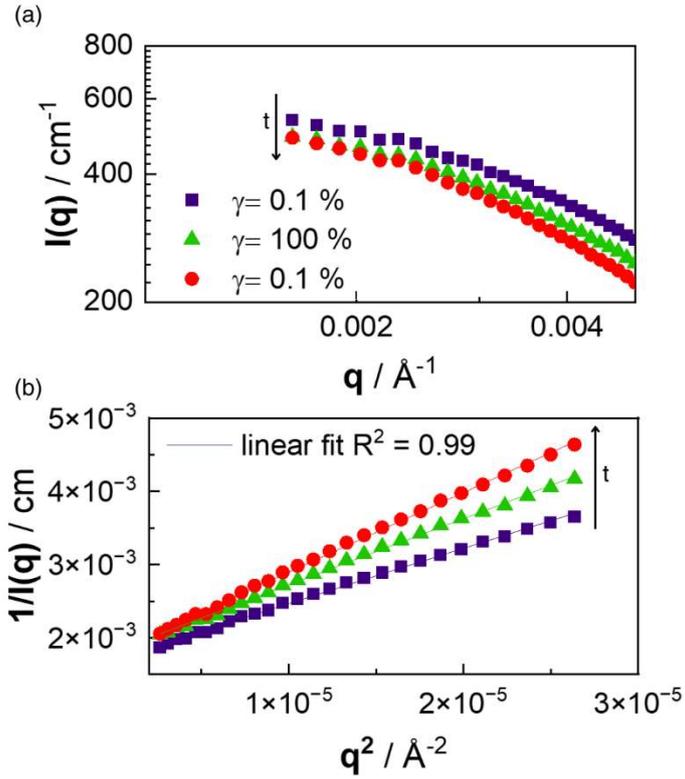

**Figure S10. (a)** Low-q region for SAXS data of G-C18:1/SCNCs hydrogel at 3 different stages of thixotropic behavior test; and **(b)** The Ornstein-Zernike plots of these profiles with their fits.

The Ornstein-Zernike equation is (Eq. 1)

$$I(q) = \frac{I(0)}{(1+q^2\xi^2)} \qquad \text{Eq. 1}$$

Which, rearranged,

$$\frac{1}{I(q)} = \frac{1}{I(0)} + q^2 \frac{\xi^2}{I(0)} \qquad \text{Eq. 2}$$

with ξ representing the mesh size of the network and I(0) is the scattering intensity at q= 0. By plotting Eq. 2 as $\frac{1}{I(q)}$ against $q^2$ (Figure S10), ξ could be estimated from the slope $\frac{\xi^2}{I(0)}$, if I(0) is known. However, this is rarely the case, as many gel and colloidal systems do not exhibit a scattering plateau at q= 0. As none of the SAXS profiles recorded in this study meet this criterion, only a qualitative approach should be considered. Since the G-C18:1/SCNCs curve is nearly flat in the low-q region (Figure S9), the true value could be obtained by extrapolation to q(0). However, for simplicity, we will use the data point measured at the lowest q-value. Table S2 present the Ornstein–Zernike plots and corresponding $\frac{\xi^2}{I(0)}$ values for the interpenetrated G-C18:1/SCNCs hydrogel at three different phases of the experiment.

Table S2. The mesh size (ξ) calculation obtained from Eq. 2 and as the slope of the Ornstein-Zernike from Figure S1010b

| Strain sweep / % | $\overline{I(0)}$ / cm$^{-1}$ | $\overline{\xi^2}/I(0)$ / Å$^2$ × cm | ξ / Å |
|---|---|---|---|
| 0.1 | 536.41 | 75.45 | 201.18 |
| 100 | 489.40 | 90.64 | 210.61 |
| 0.1 | 486.59 | 108.68 | 230.15 |

**References**

1. Poirier A, Le Griel P, Hoffmann I, et al. Ca2+ and Ag+ orient low-molecular weight amphiphile self-assembly into "nano-fishnet" fibrillar hydrogels with unusual β-sheet-like raft domains. *Soft Matter*. 2023;19(3):378-393. doi:10.1039/D2SM01218A

2. Poirier A, Le Griel P, Perez J, Baccile N. Cation-Induced Fibrillation of Micro-bial Glycolipid Biosurfactant Probed by Ion-Resolved In Situ SAXS. *J Phys Chem B*. 2022:126. doi:10.1021/acs.jpcb.2c03739ï

3. Baccile N, Poirier A, Seyrig C, et al. Chameleonic amphiphile: The unique multiple self-assembly properties of a natural glycolipid in excess of water. *J Colloid Interface Sci*. 2023;630:404-415. doi:10.1016/j.jcis.2022.07.130

4. Cui T, Tang Y, Zhao M, Hu Y, Jin M, Long X. Preparing Biosurfactant Glucolipids from Crude Sophorolipids via Chemical Modifications and Their Potential Application in the Food Industry. *J Agric Food Chem*. 2023;71(6):2964-2974. doi:10.1021/acs.jafc.2c06066

5. Baccile N, Van Renterghem L, Le Griel P, et al. Bio-based glyco-bolaamphiphile forms a temperature-responsive hydrogel with tunable elastic properties. *Soft Matter*. 2018;14(38):7859-7872. doi:10.1039/C8SM01167B

6. Almdal K, Dyre J, Hvidt S, Kramer O. Towards a phenomenological definition of the term 'gel.' *Polym Gels Networks*. 1993;1(1):5-17. doi:10.1016/0966-7822(93)90020-I

7. Raghavan SR, Douglas JF. The conundrum of gel formation by molecular nanofibers, wormlike micelles, and filamentous proteins: gelation without cross-links? *Soft Matter*. 2012;8(33):8539. doi:10.1039/c2sm25107h

8. Ben Messaoud G, Le Griel P, Hermida-Merino D, et al. pH-Controlled Self-Assembled Fibrillar Network Hydrogels: Evidence of Kinetic Control of the Mechanical Properties. *Chem Mater*. 2019;31(13):4817-4830. doi:10.1021/acs.chemmater.9b01230

9. Kantaria S, Rees GD, Lawrence MJ. Gelatin-stabilised microemulsion-based organogels: rheology and application in iontophoretic transdermal drug delivery. *J Controlled Release*. 1999;60(2-3):355-365. doi:10.1016/S0168-3659(99)00092-9

10. Phi. Thuy-Linh, Xu. W., Pernot. P., Baccile N., Kontturi E. Enhancing aqueous dispersibility of uncharged cellulose through biosurfactant adsorption. *ACS Appl Polym Mater*. Published online 2025.

11. Grachev V, Deschaume O, Lang PR, Lettinga MP, Bartic C, Thielemans W. Dimensions of Cellulose Nanocrystals from Cotton and Bacterial Cellulose: Comparison of Microscopy and Scattering Techniques. *Nanomaterials*. 2024;14(5):455. doi:10.3390/nano14050455

12. Glatter O. *Small Angle X-Ray Scattering*. Academic Press; 1982.

13. Poirier A, Le Griel P, Perez J, Hermida-Merino D, Pernot P, Baccile N. Metallogels from a Glycolipid Biosurfactant. *ACS Sustain Chem Eng*. 2022;10(50):16503-16515. doi:10.1021/acssuschemeng.2c01860

14. Teixeira J. Small-angle scattering by fractal systems. *J Appl Crystallogr*. 1988;21(6):781-785. doi:10.1107/S0021889888000263

15. Pääkkönen T, Spiliopoulos P, Knuts A, et al. From vapour to gas: Optimising cellulose degradation with gaseous HCl. *React Chem Eng*. 2018;3(3):312-318. doi:10.1039/c7re00215g

16. Mezger TG. *The Rheology Handbook*. 3rd Revised Edition. European Coatings Tech Files; 2011.

17. Seyrig C, Poirier A, Perez J, Bizien T, Baccile N. Interpenetrated Biosurfactant–Biopolymer Orthogonal Hydrogels: The Biosurfactant's Phase Controls the Hydrogel's Mechanics. *Biomacromolecules*. 2023;24(1):33-42. doi:10.1021/acs.biomac.2c00319